# Unlocking the Origin of Compositional Fluctuations in InGaN Light Emitting Diodes

Tara P. Mishra,[1, 2#] Govindo J. Syaranamual[2#], Zeyu Deng[1*], Jing Yang Chung[1,2], Li Zhang,[2] Sarah A Goodman[3], Lewys Jones[4,5], Michel Bosman[1], Silvija Gradečak,[1,2,3*] Stephen J. Pennycook[1,2*] and Pieremanuele Canepa[1,2*]

[1]Department of Materials Science and Engineering, National University of Singapore, 9 Engineering Drive 1, 117575 Singapore, Singapore

[2]Singapore-MIT Alliance for Research and Technology, 1 CREATE Way, #10-01 CREATE Tower, Singapore 138602, Singapore

[3]Massachusetts Institute of Technology, Department of Materials Science and Engineering, Cambridge, Massachusetts, 02139, USA

[4]School of Physics, Trinity College Dublin, Dublin 2, D02 PN40, Ireland

[5]Advanced Microscopy Laboratory, Centre for Research on Adaptive Nanostructures & Nanodevices (CRANN), Dublin 2, Ireland

*Corresponding authors: pcanepa@nus.edu.sg, stephen.pennycook@cantab.net, gradecak@nus.edu.sg and msedz@nus.edu.sg

**Abstract:**

The accurate determination of the compositional fluctuations is pivotal in understanding their role in the reduction of efficiency in high indium content $In_xGa_{1-x}N$ light-emitting diodes, the origin of which is still poorly understood. Here we have combined electron energy loss spectroscopy (EELS) imaging at sub-nanometer resolution with multiscale computational models to obtain a statistical distribution of the compositional fluctuations in $In_xGa_{1-x}N$ quantum wells (QWs). Employing a multiscale computational model, we show the tendency of intrinsic compositional fluctuation in $In_xGa_{1-x}N$ QWs at different Indium concentration and in the presence of strain. We have developed a systematic formalism based on the autonomous detection of compositional fluctuation in observed and simulated EELS maps. We have shown a direct comparison between the computationally predicted and experimentally observed compositional fluctuations. We have found that although a random alloy model captures the distribution of compositional fluctuations in relatively low In (~ 18%) content $In_xGa_{1-x}N$ QWs, there exists a striking deviation from the model in higher In content (≥ 24%) QWs. Our results highlight a distinct behavior in carrier localization driven by compositional fluctuations in the low and high In-content InGaN QWs, which would ultimately affect the performance of LEDs. Furthermore, our robust computational and atomic characterization method can be widely applied to study materials in which nanoscale compositional fluctuations play a significant role on the material performance.

**Introduction:**

Indium gallium nitride-based (In$_x$Ga$_{1-x}$N/GaN) light-emitting diodes (LEDs) heralded a revolution in the era of solid-state lighting with unprecedented high efficiencies and low costs[1,2]. Owing to the large tunability of their band gaps, In$_x$Ga$_{1-x}$N/GaN-based LEDs have the potential to cover the entire visible spectrum by varying the In concentration[3]. The last decade has seen an enormous commercial success of these LEDs, albeit their experimental internal quantum efficiencies (IQE) remain considerably lower than the theoretical predictions at high In concentrations[4].

Although blue commercial InGaN LEDs emitting at wavelength of ~450 nm display high IQE of ~90% and approach nearly theoretical values [5–7], the IQE values drop significantly to <60% in the longer wavelength region at >470 nm; this phenomenon is termed the "green-gap effect" [8,9]. Thus, it is imperative to unravel the origins of the green-gap to achieve the next generation of phosphor-free, longer-wavelength solid-state lighting.

To date, three main hypotheses have been put forward to understand the cause of the green-gap effect in InGaN-based LEDs. The first one refers to the possible degradation of material quality caused by the incorporation of larger concentrations of In, owing to the difference in the InN and GaN bond lengths. However, this is unlikely since in commercial LEDs the quality of these thin films does not vary appreciably in the spectral range of 400-515 nm, as experimentally determined from the Shockley-Read-Hall (SRH) recombination constants[2,10]. Further, it has been suggested that increasing the In concentration amplifies the polarization field in the (0001) polar quantum wells (QWs) [11], which could promote separation of electrons and holes, thus lowering the efficiency of radiative recombination. Current experimental evidences remain insufficient to justify the enhanced efficiencies in non-polar and semi-polar

QWs[12–15]. However, recent studies on using an (Al,Ga)N capping layer which increases the polarization field has been shown increase the IQEs at longer wavelengths[16]. Finally, the carrier localization due to compositional fluctuations and structural inhomogeneities in the InGaN QWs have also been suggested as a possible cause for the green gap effect[12,17–19]. Compositional fluctuations lead to changes in band gap resulting in lateral fields that confine the carriers to regions of smaller band gap, electrons in the conduction band and holes in the valence band.

A number of theoretical models have been proposed to explore the atomistic origins of the green-gap effect [17,18,20–22], largely focusing in random alloy models. In contrast, several optical studies of InGaN QWs, both theoretical[23,24] and experimental,[25,26] have observed properties that deviate from the random alloy model. For example, semi-empirical studies of carrier localization by large-scale (tens of nanometers) spatial compositional fluctuations in the InGaN QWs severely overestimate the IQE[12,17]. Moreover, the computationally intensive nature of first-principles techniques has hindered attempts to directly decipher the atomic microstructures of InGaN LEDs[27,28]. Chan and Zunger[29,30] investigated the atomic microstructure of low-indium concentration InGaN alloys using a multiscale approach parametrized on high-quality first-principles calculations, but did not investigate the higher concentrations in the green-gap region[26]. Consequently, it is important to investigate the accurate length scales of compositional fluctuations in higher-concentration alloys to unravel the origins of the green-gap effect.

Previously, several approaches have been applied to investigate the atomistic nature of compositional fluctuations in InGaN LEDs. In-depth studies of atom probe tomography have provided valuable insights on the In distribution in the InGaN QWs[31–36]. Extensive atomically-resolved characterization techniques, such as scanning transmission electron microscopy (STEM) [37–41] have proved the absence of large scale (on the order of tens of nanometers) compositional fluctuations. Nevertheless, atomistic-scale investigations using high-resolution electron microscopy suffer from a limited field of view, which hinders the detection of accurate size distribution of potential compositional fluctuations[37–41], and possible beam-induced structural modifications[42]. Recently, it has been shown that imaging under the knock-on threshold preserves the QW structures[38,43], allowing the quantification of compositional fluctuations at the sub-nanometer scale. To measure the spatial distribution of compositional fluctuations, a larger area of the sample needs to be investigated with sufficiently high resolution.

Here, we use a combination of atomically resolved electron energy loss spectroscopy (EELS), and a multi-scale theoretical approach to elucidate the intrinsic compositional fluctuations of $In_xGa_{1-x}N$ alloys up to $x \approx 24\%$ In, the region of the green gap, and demonstrate their non-random nature. We obtained a large number of maps of the InGaN QWs and quantify the compositional fluctuations using an autonomous scale-space method based on the convolution of Laplacian of Gaussians (LoG)[44], which can detect the compositional fluctuations without any prior assumption. The LoG algorithm has been shown previously to very reliable and robust in autonomous feature detection[45,46]. Furthermore, we used a multiscale theoretical approach, based on density functional theory (DFT), cluster expansion (CE), Monte Carlo (MC) to predict, without any initial assumptions, the intrinsic atomic distribution in the $In_xGa_{1-x}N$ QWs. We bridge the gap between the experimentally and computationally observed compositional fluctuations by simulating STEM-EELS images from MC models through a "multislice" approach. We demonstrate an excellent match between the experimental observations of compositional fluctuations of the InGaN microstructures and the predictions by Monte Carlo. This study develops a systematic framework for understanding the origin and existence of intrinsic compositional fluctuations in InGaN QWs, which is essential for the development of the next generation of longer wavelength LEDs covering the entire visible spectrum.

**Results:**

**Characterization of the InGaN Quantum Wells:**

**Figure 1a** shows a cross-section of the LED device used in this investigation, which follows the fabrication protocol by Zhang *et al*.[47] and is discussed in the method section. The sample contained three sets of quantum wells, each set consisting of three alternating InGaN QWs and GaN barrier layers with thickness 3 nm and 10 nm, respectively.

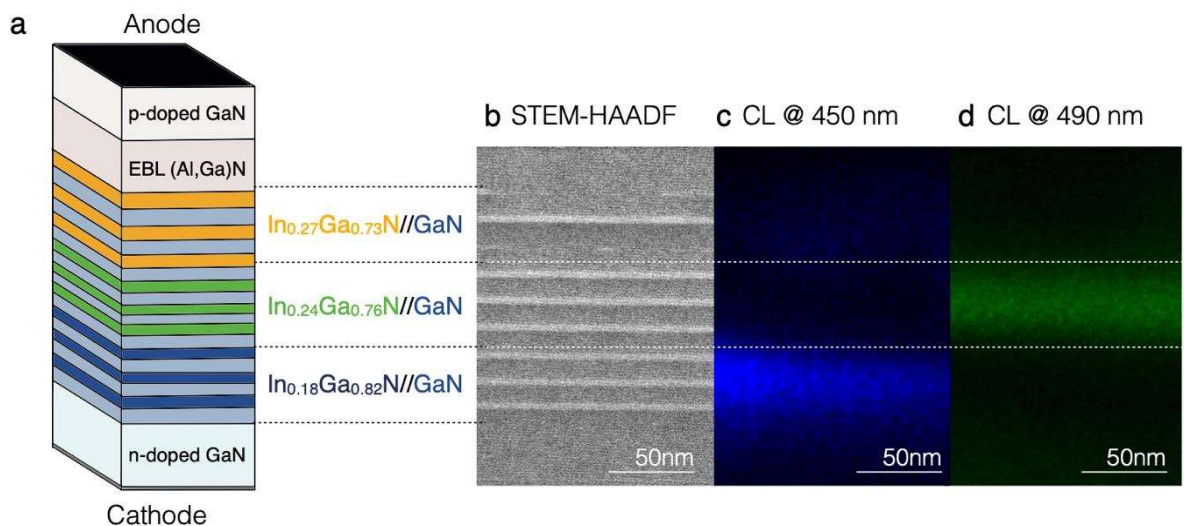

**Figure 1:** LED device and its characterization. **(a)** Cross-section of the LED device composed of alternating InGaN QWs and GaN barrier layers, **(b)** Low Magnification STEM-HAADF Image from the QW region of the LED, **(c,d)** The corresponding monochromatic STEM CL images recorded at 450 nm and 490 nm. Images have been false colored according to the CL wavelength.

The nominal In composition increased sequentially going from the bottom QW set toward the top one, which was confirmed by our measurements. A low-magnification high angle annular dark field (HAADF) STEM image (**Figure 1b**) micrograph shows three distinct QW regions mapped to the device in Figure 1(a). Layers of InGaN can be seen in **Figure 1b**, which match the bottom and middle QWs. However, the top 3 QWs with the highest indium content do not display the same structural integrity as compared to the middle and bottom QWs, and therefore these have not been considered for the analysis of the compositional fluctuations.

We further used EELS measurements to determine In content in each set of QWs. The procedure to estimate the In content of the QWs is explained in the supplementary information (SI). The nominal In composition from EELS of the bottom and middle sets of QWs are 18.4 ± 2.4% and 24.2 ± 2%, *i.e.* composition $In_{0.18}Ga_{0.82}N$ and $In_{0.24}Ga_{0.76}N$. We have performed cathodoluminescence (CL) measurements to confirm the indium content of these QWs. Monochromatic CL images obtained at emission wavelengths centered at 450 nm (**Figure 1c**) and 490 nm (**Figure 1d**) can be directly correlated to the bottom and middle sets of QWs, respectively. By correlating these emission wavelengths with predicted band-gaps of $In_xGa_{1-x}N$ from accurate hybrid functional DFT calculations by Moses *et al*,[48] we deduced In contents of ~16.5% and ~22.2% in the bottom and middle QWs (**Figure 1a**), respectively. The nominal compositions determined from the CL measurements are in good agreement with the those obtained from EELS measurements. For the remainder of the paper, we will refer to these two In concentrations as low and high, respectively.

**Autonomous Detection of Compositional Fluctuation in $In_xGa_{1-x}N$**

To quantitatively determine the length scale of In compositional fluctuations in the $In_xGa_{1-x}N$ alloys we used a multifaceted approach, which combines experimental

EELS maps and predicted maps from independent Monte Carlo simulations, as in **Figure 2**.

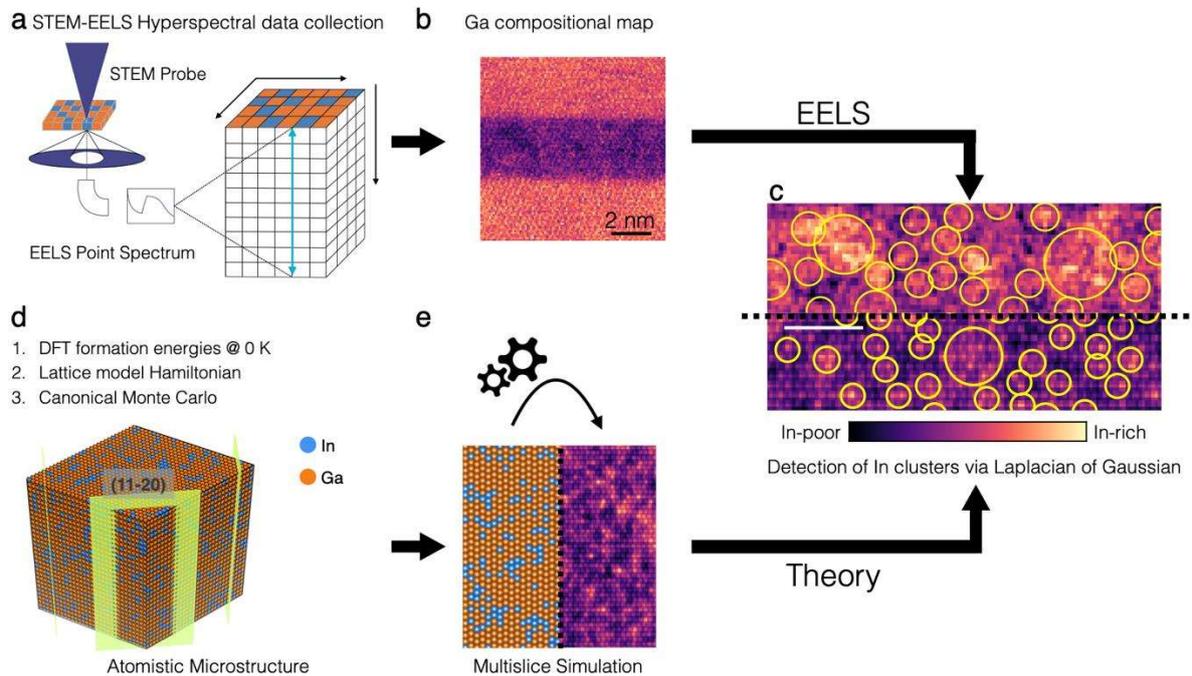

**Figure 2:** Autonomous detection of compositional fluctuations in $In_xGa_{x-1}N$ samples. Panels **(a)**, **(b)** and **(c)** show the analysis of the experimental STEM-EELS maps of the $In_xGa_{x-1}N$ samples, whereas panels **(d)**, **(e)** and **(c)** depict the generation and analysis of similar maps from a theoretical approach. Panel **(a)** depicts a schematic STEM-EELS hyperspectral measurement. Panel **(b)** shows a representative Ga compositional maps which are integrated from the hyperspectral dataset. Panel **(c)** shows the detection of In compositional fluctuation in the micrographs from STEM-EELS with the color bar representing the In concentration (top) and theory (bottom), respectively, on which a convolution of the Laplacian of Gaussian (LoG) kernels with the input image is performed. The white bar represents the 2 nm scale. Panel **(d)** shows a representative microstructure model obtained using canonical Monte Carlo **(e)** and viewed from the [11-20] axis. A multi-slice simulation of the EELS map, using the MULTEM package, from the Monte Carlo microstructure model is shown.

These two seemingly distinct strategies are independent and applied in parallel to clarify the phenomenon of In compositional fluctuation in InGaN QWs.

The EELS data were collected using a STEM probe rastering across the InGaN layers, thereby forming a hyperspectral data prism as shown in **Figure 2a**. The EELS spectra of the QWs were collected using an 80-kV accelerating voltage (**Section S1** and **Figure S1** in SI). This accelerating voltage is optimal[43] for imaging InGaN QWs by preventing the knock-off beam damage[38] for both light and heavy atoms, and at the same time it provides sufficient resolution for imaging. Under these imaging conditions, QWs remain structurally pristine compared to the fuzzy and irregular QWs generally seen due to beam damage[42].

The electron energy loss spectra were quantified using the Ga *L* edge (~ 1115 eV), as shown in **Figure 2b**. The Methods Section describes the procedure and parameters used to process the EELS spectra. We have surveyed ~ 500 nm$^2$ of QW region for each QW at two specific In concentrations (shown in **Figure 1**) to undertake a statistical analysis of the compositional fluctuation length scale.

In parallel, we undertook a novel methodology to theoretically assess the compositional fluctuations in InGaN-based QWs. While DFT has been used extensively to study the properties in InGaN system[48,49], DFT alone cannot handle extremely large models with thousands of atoms, which are much needed to unravel the alloying properties of this material. Here, we propose a multi-scale approach to reveal the compositional fluctuations in realistic samples.

To assess the spontaneity of mixing of Indium in the InGaN QWs, a large number of configurations (or orderings) describing the distribution of In into the wurzite structure of GaN (*P*6$_3$*mc*) were simulated. Eq. 1 defines the formation energy ($E_f$) of each ordering, computed by DFT, with respect to the energy of GaN and InN —the end-members.

$$E_f[In_xGa_{1-x}N] = E[In_xGa_{1-x}N] - xE[InN] - (1-x)E[GaN] \qquad (1)$$

where $E[In_xGa_{1-x}N]$, $E[InN]$ and $E[GaN]$ are the DFT total energies of the $In_xGa_{1-x}N$ orderings, and those of the reference structures of InN and GaN, respectively. We probed the incorporation of Indium in GaN by considering a number of supercell models ranging in sizes from the primitive cell (comprising only 2 formula units) of the wurzite-GaN structure to large 2x2x2 supercell models. We have also considered the 3x2x1 and the 2x1x3 supercells; the latter is important to capture the long-range interactions along the growth direction (i.e., [0 0 0 1]) of the QWs. These supercell structures are needed to finely sample the In content in In$_x$Ga$_{1-x}$N (with a step of Δ$x$ = 0.08), as well as to explore the configurational ordering of In.

In particular, to assess the strain characteristics of the InGaN QWs, we developed two distinct models: i) the bulk incoherent model where both the volume and the shape of each In$_x$Ga$_{1-x}$N ordering are fully relaxed. Thus, using DFT the GaN-InN tie-line was described by computing the mixing energies of 575 distinct In$_x$Ga$_{1-x}$N orderings. ii) The epitaxially coherent model, where the supercell models are constrained to the GaN substrate in the plane normal to the [0001] direction (the *ab* plane) but allowing the cell and volume relaxation along the [0001] direction. Note, the [0001] is the typical growth direction of the polar InGaN LEDs. Thus, the formation energies of 643 distinct epitaxially coherent orderings were computed.

The formation energies of In into GaN across the entire composition provide the phase diagrams InN–GaN at 0 K, which are reported in **Figure 3a** for the bulk incoherent and **Figure 3b** for the epitaxially coherent models, respectively.

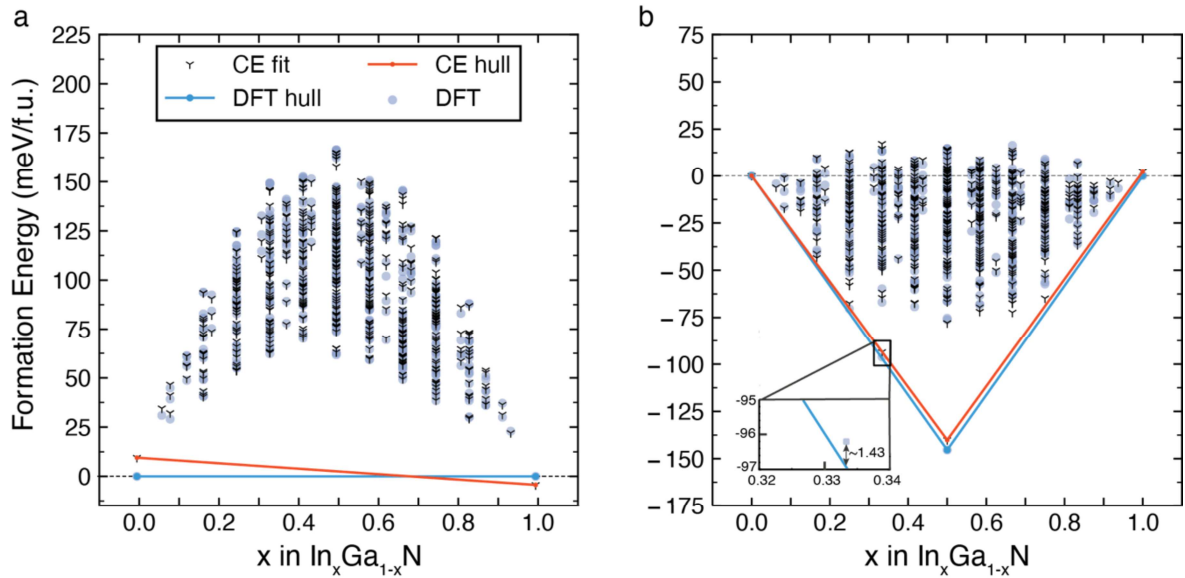

**Figure 3:** Formation energies calculated using DFT (blue dots) and the CE Hamiltonian (black cross) for InGaN QWs in **(a)** the bulk incoherent model and **(b)** the epitaxially coherent model. The respective error plots of both CE are reported in **Figure S5** in SI. Inset in panel **(b)** highlights the distance from the convex hull (~1.43 meV/f.u.) of the DFT metastable ordering at $x = \frac{1}{3}$ (In$_{1/3}$Ga$_{2/3}$N).

At a given composition, the ground state structure always minimizes the Gibbs energy at that composition. Therefore, given a set of InGaN compositions, their structures, and their DFT energies, the equilibrium phase diagram at 0K is mathematically obtained through a procedure of convex minimization. The convex-hull construction evaluates the stability of a given compound against any linear combination of compounds that have the same composition. For the bulk incoherent model of **Figure 3a**, the mixing energies remain positive at all Indium compositions, which indicates that at 0 K In cannot favorably mix with GaN to form In$_x$Ga$_{1-x}$N.[29,30]

The picture changes entirely when the QW undergoes epitaxially coherent strain, with a stable ordering located at composition x = 0.50 (In$_{0.5}$Ga$_{0.5}$N). Furthermore, a slightly metastable ordering at $x = \frac{1}{3}$ (In$_{1/3}$Ga$_{2/3}$N) is observed, with a very small energy above the convex hull (~1.43 meV/f.u). The slightly metastable and stable compositions at x = $\frac{1}{3}$ and x = 0.5 can be mapped to well-defined ordered arrangements, which are the $(InN)_1/(GaN)_2$ and $(InN)_2/(GaN)_2$, respectively and are shown in **Figure S4** of the SI. The ordered structure $(InN)_2/(GaN)_2$ matches with previous predictions by Liu and Zunger[24]. One important conclusion form **Figure 3b** is that in the epitaxially coherent model the phase separation will take place at 0 K for Indium composition in the range 0 < x < 0.5 and 0.5 < x < 1.

Using the DFT mixing energies of the incoherent and epitaxially coherent InGaN orderings at 0 K, we developed a cluster expansion (CE) Hamiltonian based on the approach by Sanchez et al[50], whose details and derivation are described in the method section.

Having developed a robust CE model, that directly accounts for the effects of In incorporation, the local environment and strain on the mixing enthalpies, we employ Grand Canonical Monte Carlo[51] (GCMC) to predict the In composition *vs.* temperature phase diagrams for the InGaN alloys for the bulk incoherent model (**Figure 4a**) and the epitaxially coherent case (**Figure 4b**).

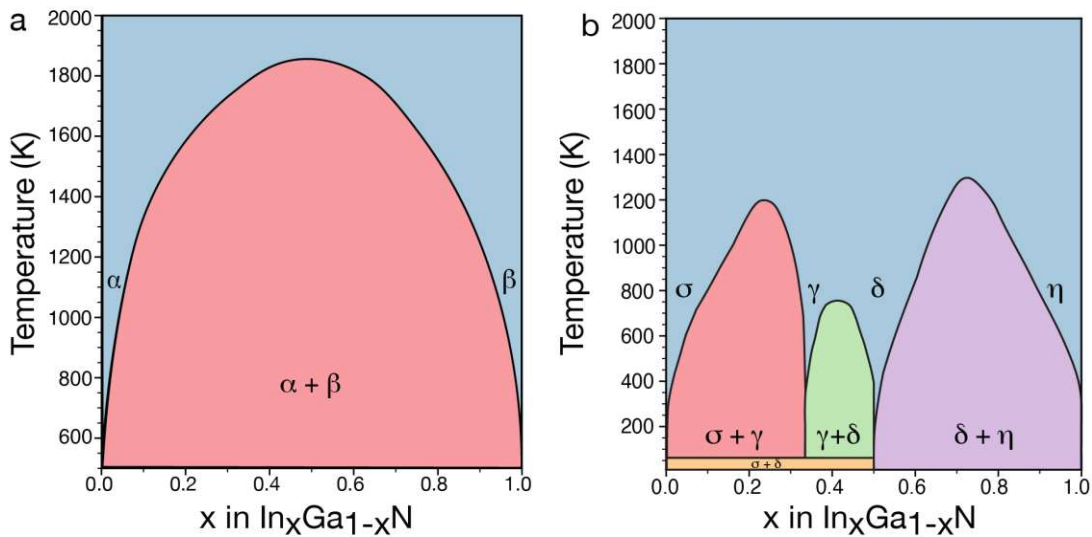

**Figure 4:** Compositional phase diagram of the $In_xGa_{1-x}N$ alloy as a function of temperature as derived from the GCMC simulations parameterised on the CE model for **(a)** the bulk incoherent

and **(b)** the epitaxially coherent InGaN QW model. In the bulk incoherent model **(a)**, the red area ($\alpha + \beta$) represents the biphasic region, while the blue area ($\alpha$ and $\beta$) shows the single phase regions. A miscibility gap is observed till a critical temperature of 1950 K. In the biphasic region the alloy has the tendency to phase separate into Ga ($\alpha$) and In ($\beta$) rich regions. For the epitaxially coherent model of panel **(b)**, three stable phases are observed at ~0 K, namely, the Ga rich phase ($\sigma$), the (InN)$_2$/(GaN)$_2$ ordered phase ($\delta$) and the In rich phase ($\eta$). Above ~100 K, at x = $\frac{1}{3}$ a new single phase ($\gamma$) is observed. The phase $\gamma$ decomposes into phases $\sigma$ and $\delta$ through a eutectoid transformation at ~80 K. The single phases are always separated by biphasic regions, while full In solubility is obtained for temperatures above ~1300 K.

In the bulk incoherent model, the InGaN system exhibits phase separation for the complete composition range. Above the critical temperature of 1950 K, Indium becomes fully miscible in GaN. This is in excellent agreement with previous reports, [29,30] and provides credibility to our theoretical model.

In contrast, the predicted phase diagram of the epitaxially coherent case presents a monophasic phase ($\delta$) at 0 K at composition of ~ 50 % along with the Ga rich phase ($\sigma$) and In rich phase ($\eta$). A new monophasic region ($\gamma$) at x ~ 33.33% is observed beyond ~80 K through an eutectoid transformation between the phases $\sigma$ and $\delta$ (**Figure 4b**). Above 80 K, the $\sigma + \gamma$ biphasic region (red in **Figure 4b**) covers a large portion of the technologically relevant range of In concentration (0 < x < $\frac{1}{3}$) and the full miscibility of In in GaN is only accessible at ~ 1100 K and x ~ 24 %. In **Figure 4b**, two other biphasic regions at higher In concentrations are observed: i) $\gamma + \delta$ in the range $\frac{1}{3}$< x < 0.5, and ii) $\delta + \eta$ in the range 0.5 < x < 1.

To gain further understanding of the microstructure of the InGaN QWs at specific In concentrations and temperatures, we performed canonical Monte Carlo (CMC) simulations based on the CE approach developed for the bulk incoherent and the epitaxially coherent models. A representative snapshot of an In$_x$Ga$_{1-x}$N supercell model from CMC is shown in **Figure 2d**. To obtain a meaningful comparison of the CMC predictions with the In distribution from the EELS maps, the predicted structures were oriented along the direction of imaging of the STEM-EELS, *i.e.* the [11-20] zone axis **Figure 2b**. A simulation to introduce the incoherent scattering as observed in the EELS compositional maps was performed using the MULTEM package (see details in the Method Section)[52,53]. **Figure 2b** shows a simulated final map obtained from the CMC calculations and the EELS simulation.

To quantitatively determine the spatial extent of In aggregation in In$_x$Ga$_{1-x}$N QWs from the EELS and the theoretically CMC predicted maps, respectively, we have developed an autonomous strategy to detect compositional fluctuations, whereby the maps produced undergo a convolution with kernels of Laplacian of Gaussians (LoG). The LoG algorithm convolutes the input image, either from EELS or CMC simulations, with a series of LoG kernels of discrete variances ($\sigma$) to find local compositional fluctuations in the QWs. The LoG kernels can be extended to detect anisotropy in the compositional fluctuations as shown in section S9 in the supplementary information. However, for simplicity we have assumed circular Guassian kernels to map the compositional fluctuations. The conditions used to perform the LoG convolutions are described in the Method Section. A statistical analysis using the LoG kernels was performed and the average size of In compositional fluctuations were identified. **Figure 2c** shows an example of EELS and CMC maps on which the LoG strategy is applied; compositional fluctuations are identified by yellow circles of different diameter.

**Distribution of Compositional Fluctuations from Experiments and Theory:**
We now turn to an analysis to determine if the observed compositional fluctuations can be described using the random alloy model or indicate some additional clustering.

The LoG detection method was applied to 17 experimental EELS maps for each of the two In content QWs, as well as the simulated images (15 images **Table S3**) for each CMC temperature (see the Method Section). **Figures 5a** and **5b** show representative EELS micrographs from the low- and high-indium content QWs analyzed using the LoG algorithm. The compositional fluctuations are highlighted as yellow circles. The diameter of the circles was calculated from the obtained variance ($\sigma$) of the gaussian kernels used in the LoG algorithm (Eq. 6).

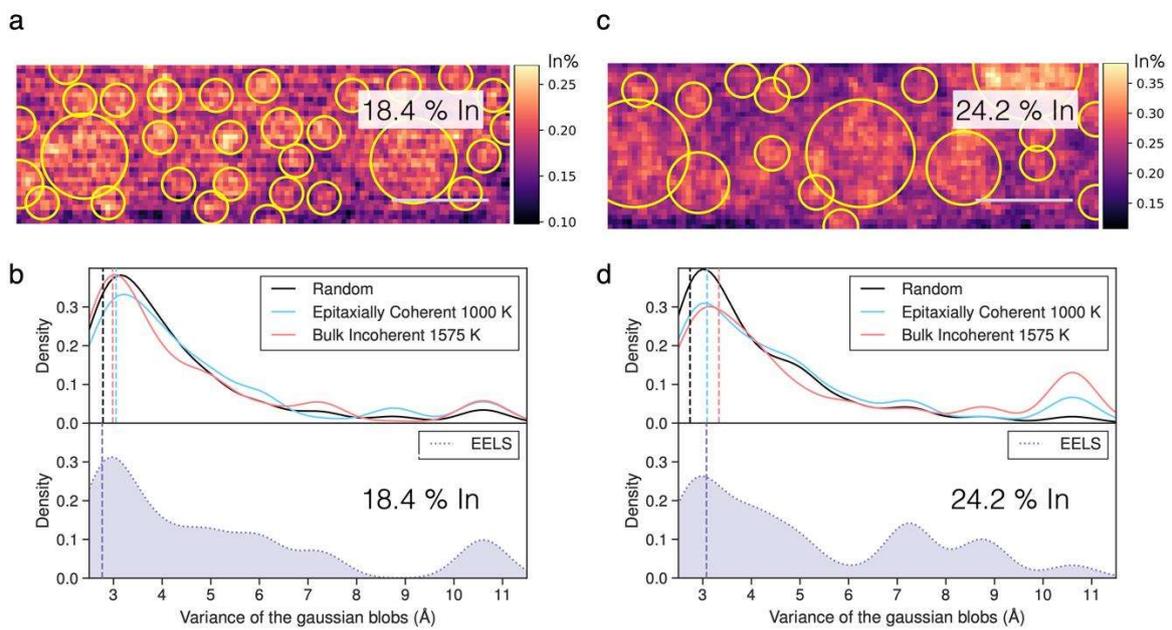

**Figure 5:** Representative EELS compositional maps of QW regions overlaid with circles denoting compositional fluctuations found using the LoG algorithm for QWs with In composition **(a)** 18.4% and **(c)** 24.2%. The scale bars in **(a)** and **(c)** are 2 nm and shown by grey bars. The probability density functions (y-axis) of the size of the compositional fluctuations (x-axis) in **(b)** 18.4% and **(d)** 24.2% In content QWs, respectively are obtained from EELS experiments (in violet, bottom) and compared with different canonical Monte Carlo simulations (CMC) at conditions namely, random at 4000 K (black), epitaxially coherent at 1000 K (blue) and bulk incoherent at 1575 K (red). The random alloy distribution is obtained by running the CMC simulations at a temperature of ~4,000 K, which is significantly higher than the miscibility temperature of In in GaN (see **Figure 4a** and **Figure 4b**). The average size of the compositional fluctuation is marked by dashed lines.

From the EELS maps, the LoG strategy detects significantly smaller compositional fluctuations in the QW with low In compared to QWs with higher In content. For example, comparing **Figures 5a** and **5c**, one can immediately observe that the compositional fluctuations are larger in the system with higher concentration. We use a similar analysis over a large data set of micrographs and predicted EELS pictures to inform the nature of the In compositional fluctuations.

The compositional fluctuations are treated as connected pixels with higher intensity these are alternatively referred to as blobs (Binary Large Objects). We now analyze the In compositional fluctuations (blobs) revealed by low and high In content InGaN QWs as shown in **Figures 5b** and **5d**. These figures show the probability densities (*y*-axis) estimated via the kernel density, as a function of variance of the gaussian blobs (*x*-axis). The variance of the gaussian blobs represents the distribution of indium compositional fluctuations in the QWs. The size of In compositional fluctuations were revealed experimentally using EELS (violet, in the bottom panel of **Figures 5b** and **5d**). The probability density functions in the top panels of **Figures 5b** and **5d** show the size of the computational fluctuations derived theoretically from CMC models (black, blue, red) at selected temperatures. These are: i) 4000 K mimicking the random arrangement, ii) the epitaxially coherent at 1000 K, and iii) the bulk incoherent at 1575 K. The average sizes of the compositional fluctuations obtained from the experimental and theoretical calculations are marked by the dashed vertical lines. The distribution of compositional fluctuations analyzed for all the CMC temperatures are summarized in **Figure S6** and **S7** for bulk incoherent and bulk epitaxially coherent InGaN QWs, respectively.

The LoG analysis of EELS maps applied to the low In content QWs (**Figure 5c**) shows that most (~ 90%) of the compositional fluctuations are less than ~4 Å in variance (~ 9.42 Å FWHM), with only a small fraction (~ 9.5 %) of fluctuations greater than ~4 Å. In **Figure 5d** we observe that the high-In content QWs show significantly larger probability density (~ 17 %) of blob sizes greater than ~4 Å compared to the low-In content LEDs. In this manuscript we have compared fluctuations larger than 4 Å in variance as large-scale compositional fluctuations since they correspond to ~ 10 Å full width half maxima (FWHM).

Care is needed in comparing the EELS data and the CMC simulation because the InGaN/GaN LEDs are grown through epitaxy, a situation that is far from thermodynamic equilibrium[54]. In contrast, our CMC simulations predict properties at thermodynamic equilibrium, and thus a one-to-one comparison with the fabrication temperatures of the LED devices is not appropriate. Therefore, the Monte Carlo calculations were performed over a range of temperatures for both bulk incoherent (1350 K to 4000 K) and epitaxially coherent (500 K to 4000 K) models to determine the In distribution in the InGaN QWs.

The variation of the average compositional fluctuations ($x$-axis) with the CMC temperatures ($y$-axis) is shown in **Figure 5** and compared to the EELS measurements. In general, it is expected that as the temperature of the system increases, the solubility of In in GaN increases proportionally (see phase diagrams in **Figure 4a** and **Figure 4b**). The average sizes of the compositional fluctuations obtained from CMC simulations at different temperatures confirm this behavior. The blob sizes of **Figure 5** display that the tendency for In fluctuations decreases monotonically as the temperature of the simulation is increased. CMC simulations at exceedingly high temperatures, *i.e.,* 4000 K (indicated as random in **Figures 5c** and **5d**) are used as an extreme case to verify the full solubility of In into GaN. Any compositional fluctuation detected in such QWs reflect purely statistical fluctuations, hence, these high-temperature structures serve as a reference to describe a fully random distribution of indium in InGaN QWs. Notably, CMC simulations based on lattice-type models forbid the transition of $In_xGa_{1-x}N$ system to a liquid, but in contrast they are appropriate to capture the regime of a random alloy. In the fully random limit, using the LoG analysis we show the absence of compositional fluctuations of appreciable size (> 4 Å in variance for both 18.4% and 24.2%) in the QWs (**Figure 5b** and **5d**).

The appropriate temperature of representative CMC simulations was identified by performing a matching of the average blob size obtained at several temperatures with the EELS data. The temperature ranges inspected are: i) bulk incoherent from 1375 K to 4000 K, and ii) for the epitaxially coherent 500 K to 4000 K. From **Figure 5c** it is observed that the best match between EELS and CMC images of low In content $In_xGa_{1-x}N$ QWs is achieve at ~4000 K, which is the random alloy model. However, for the high In content $In_xGa_{1-x}N$ QWs, as seen in **Figure 5d**, the EELS distribution are best matched at ~1575 K for the bulk incoherent model and at ~1000K for the epitaxially coherent model. Therefore, the independent computational approach confirms the EELS observations by showing an increase in the large-size compositional fluctuations (> 4 Å in variance), accounting for ~18 % in high In QWs. As expected, the low In content QWs shows a smaller fraction (~10.5 %) of large-size compositional fluctuations. Notably, the description of the compositional fluctuations by CMC in the epitaxially coherent QWs is best described by lower temperature models (~1000 K) compared to the bulk incoherent system found at higher temperatures (~1575 K). This temperature decrease can be linked to the bi-axial strain experienced by the InGaN QW, which are typically grown at ~ 1000 K.[47]

The intrinsic tendency of the InGaN QWs for compositional fluctuations is also reflected by the computational data. For all temperatures the simulations consistently predict a larger length scale compositional fluctuation in the higher indium content LEDs as shown in **Figure 5, S5 and S6**, compared to those with lower indium concentration.

**Discussion:**

In this manuscript, we investigate the presence of In compositional fluctuations in 450 nm ("low In content") and 490 nm ("high In content") emitting InGaN-based LED utilizing a multifaceted methodology combining materials and device fabrications, STEM-EELS and CL together with a novel multi-scale modeling approach. We provide a statistical assessment of the LED QWs, from which we derive the distribution of In fluctuations in the InGaN/GaN QWs. We carried our experiments at non-damaging conditions to the QWs[38] and have provide evidence for the same.

Using these STEM-EELS micrographs we present a quantitative analysis of the size of In compositional fluctuations in InGaN LEDs. We developed a novel methodology based on LoG that we applied to the STEM-EELS experimental and simulated micrographs to detect indium compositional fluctuations in the InGaN QWs. The novel LoG method developed in the study removes the bias of initial position guesses of suspected In fluctuations and eliminates the need for arduous pre-processing of the EELS micrographs. This technique is of general applicability and can be adapted to any materials science problems study where compositional fluctuations need to be investigated.

Here, we have predicted phase diagrams for the bulk incoherent and epitaxially coherent InGaN QWs. In the bulk incoherent system, we could observe that InN has a strong tendency to phase separate from the GaN lattice. The same observation can be extended for the epitaxially coherent InGaN QWs for In composition in the range $0 < x < \frac{1}{3}$.

Our predicted phase diagram of **Figure 4b** shows two dominant monophasic regions stable above ~80 K and at ~ 33.33 % ($\gamma$) and ~ 50 % ($\delta$) which are separated by biphasic regions from 0 < x < $\frac{1}{3}$ and $\frac{1}{3}$< x < 0.50. Importantly, these findings support the idea that even in the epitaxially coherent models, Indium is not miscible in GaN up to 33%. The epitaxial coherent model shows the lowering of the critical temperature (~ 1200 K, x ~ 24 % **Figure 4b**) beyond which In can always mix with GaN compared to the bulk incoherent case (~ 1500 x ~ 50 %, **Figure 4a**). Therefore, for the high In content LEDs (x ~ 24.2 %), the growth temperatures (~ 1000 K) corresponds to the biphasic region, which results in phase separation. However, for the low In content (x ~ 18.4 %), InGaN LEDs grown at higher temperatures of ~ 1033 K corresponds to the boundary between the biphasic ($\sigma + \gamma$) and the Ga rich monophasic ($\sigma$) region in the epitaxially coherent phase diagram (**Figure 4b)**. Hence, in the low In composition LEDs, phase separation is suppressed.

Although the seminal phase diagram proposed by Karpov predicts In solubility in the range 0 < x < 35 % and at low temperatures, our findings remain in striking contrast [55]. Furthermore, we showed that the suppression of the phase separation in epitaxially coherent systems (compared to the bulk incoherent system) is driven by a reduction of the critical miscibility temperature for the low In content InGaN QWs. This is accompanied by the stabilization of intermediate ordered phases at x ~ 33.33 % and ~ 50 %. Our prediction of the phase segregation behavior in the very high In (~ 82%) content InGaN QWs is in great agreement with the previous experimentally reported [56] and first-principles based phase separation study[30,57]. Figure 4b shows that at temperature of ~1000 K InXGa1-xN with In content from 15 % to 30 % should phase separate and furthermore 35 % to 60 % should not phase separate ($\gamma$ phase). However, In concentrations above 35 % are hard to grow and can show precipitation of In rich droplets[58]. This is not predicted in our current multiscale thermodynamic model as it does not account for all the kinetic and metastable effects during the growth of the InGaN QWs.

Here, we have developed large-scale CMC models (parametrized on accurate DFT calculations) with varying indium concentrations (18.4% and 24.2%). The microstructure of InGaN QWs is largely controlled by the kinetics imposed by their epitaxial growth, which probably limits the observation of a complete phase separation into InN and GaN, as predicted by our 0 K phase diagram from DFT. To this end, we identified the representative CMC temperature conditions describing the metastable situation imposed by the constrained growth of InGaN LEDs and simulated the microstructures with Monte Carlo.

From our EELS and computational models, we find a nearly two-fold increase in nanometer scale (> 1 nm in FWHM) compositional fluctuations for the high In content QWs (~ 24.2 % In) compared to the low In content QWs (~ 18.4 % In).

In low In content LEDs $In_{0.15}Ga_{0.85}N$, Chichibu *et al.*[26] have observed prominent localization of charge carriers and attributed them primarily to the phenomenon of In compositional fluctuations. Our findings underline that larger-scale compositional fluctuations are present in high InGaN LEDs (In content ~ 24%) and may play a bigger role in the performance of the LEDs. These independent findings are in excellent agreement with the EELS-STEM observations as well as CMC predictions and confirm unequivocally the existence of substantial In fluctuations InGaN LEDs. Furthermore, as seen in the epitaxially coherent case we present the separation into a GaN rich region and a $(InN)_1/(GaN)_2$ single-phase region. These atomic scale microstructures would have a significant effect on the optical properties of InGaN QWs at higher In content.

In the fully random limit, we show that Monte Carlo simulations performed on both In$_{0.17}$Ga$_{0.83}$N and In$_{0.22}$Ga$_{0.78}$N (of **Figure 5c** and **d**) at very high temperature and coupled with the LoG analysis show compositional fluctuations of appreciable size in the QWs are nearly absent.

The distribution of compositional fluctuation of low-In (**Figure 5c**) content LEDs are similar to the fully miscible solid solution with a small fraction of nanometer scale fluctuations. This observation is in agreement with the previously reported atom probe tomography measurements (APT) on InGaN QWs of similar In content [32,34]. On the other hand, in the high-indium content QWs (**Figures 5d**, **S6** and **S7**) there exists a significant difference between a complete solid solution at 4000K (random fluctuation model) and the best match distribution at ~1000 K for the epitaxially coherent model. This CMC temperature is in good agreement with the growth temperature of the high In content LEDs. Consequently, the epitaxially coherent model is able to predict the In distribution in the InGaN QWs across different In concentrations. The minor deviations between the CMC temperatures and the experiment temperature could be attributed to the surface effects during the growth process, which influence the Indium incorporation in the InGaN QWs[59,60].

In the high In content LEDs a partial phase separation is observed where random compositional fluctuations are concurrent with pockets of In rich regions. Our observations are in good agreement with the previously observed compositional fluctuations using APT on InGaN QWs of high In content (~ 25 % In)[31,35,36]. Based on these findings, we can assume that the currently widely accepted model of a perfect random structure does not fully capture the complete compositional fluctuations of the InGaN QWs. To date, the random alloy model of InGaN QWs is often assumed in the literature to explain the green-gap effect in high In content LEDs[18,21,31,34,61].

Furthermore, the approach provided here establishes a systematic pathway to capture the compositional fluctuations for a range of In concentrations in the InGaN LEDs including the green-gap region.

**Conclusions**

In summary, using a multifaceted methodology, we have investigated the complete distribution of the compositional fluctuations at sub-nanometer length scales in $In_xGa_{1-x}N$, LED devices. Our results suggest compositional fluctuations predominantly in the range of random fluctuations in the lower indium content. In striking contrast, we identify the occurrence of non-random indium fluctuations for In-rich QWs. These quantitative measurements pave the avenue for more investigations to elucidate the role of compositional fluctuations in the new generation of InGaN LEDs and toward practical design strategies to curb the degradation of these devices. While we have developed and applied a novel methodology, based on the convolution of Laplacian of Gaussian to rationalize the microstructure of InGaN LEDs, this strategy can be extended to other problems in materials science where understanding the compositional fluctuations is crucial.

**Methods:**

**First-principles Calculations, Cluster Expansion, and Monte Carlo Simulations:**
The incorporation of In into GaN was modeled using the cluster expansion (CE) [50] method where the mixing energies (see **Eq. 1** and **Eq. 2**) are obtained from DFT. The CE model whose details are given in Section S5 in SI was fitted using the compressive sensing algorithm [62] as implemented in the cluster assisted statistical mechanics (CASM) code[63–66].

DFT[67,68] employing the sufficiently constrained and appropriately normed (SCAN) meta-GGA exchange correlation functional[69] was performed using a combination of plane-waves and projector augmented-wave (PAW) potentials to describe the InGaN wave-functions, as implemented in VASP.[70,71] In the PAWs the following electrons were treated explicitly: In $4d^{10}5s^25p^1$ (06Sep2000), Ga $3d^{10}4s^24p^1$ (06Jul2010) and N $2s^22p^3$ (08Apr2002). A plane-wave basis with an energy cut-off of 520 eV and a 8x8x5 Gamma centered Monkhorst-Pack $k$-point mesh[72,73] were used for the primitive cell of GaN (with 2 formula units and $P6_3mc$ space group). As for the larger supercells, the $k$-point meshes were adjusted to obtain the same sampling of the 1st Brillouin zone. The total energy was converged within $10^{-5}$ eV/cell and the interatomic forces (and stresses) were converged to less than $10^{-2}$ eV/Å (0.29 GPa).

We used the formation energies evaluated with DFT to construct the cluster expansion (CE). The CE is written as a truncated summation of the effective cluster interactions (ECIs) of pair, triplet and quadruplet terms according to Eq. 2.

$$E_f(\sigma) = \sum_\alpha V_\alpha \phi_\alpha(\sigma) \qquad (2)$$

where $E_f$ is the DFT formation energy, $V_\alpha$ is the effective cluster interactions (ECIs) of a cluster $\alpha$ including its multiplicity (symmetry). $\phi_\alpha(\sigma)$ is the cluster function of cluster $\alpha$ for the configuration of possible orderings $\sigma$, which assumes the values of 1 and 0 for In and Ga respectively. Thus, the non-zero ECIs represent only geometrical clusters fully constituted by In.

The $\phi_\alpha(\sigma)$ were generated within a maximum radius extending to 12 Å, 7 Å and 6 Å for the pairs, triplets and quadruplets, respectively. The identification of only the descriptive ECIs leveraged the compressive sensing algorithm to penalize the L1-norm of the ECIs and thus minimize overfitting, which employed $\alpha$ of $1\text{x}10^{-4}$ and $3.5\text{x}10^{-5}$ for the bulk incoherent (**Figure 3a**) and epitaxially coherent models (**Figure 3b**).

Following this approach for the bulk incoherent QW, a CE of model based on 36 ECIs with a root mean square error (RMSE) of ~ 2.53 meV/f.u. and leave one-out cross-validation (LOOCV) score ~ 4.57 meV/f.u was developed. For the epitaxially coherent QW, we developed a CE model using 41 ECIs with an RMSE ~ 3.04 meV/f.u. and a LOOCV score ~8.04 meV/f.u.

The CE is then used to express the free energy of large-model supercells in two types of Monte Carlo simulations, i.e. the grand canonical (GCMC)[51] ensemble to study the phase diagram and the canonical Monte Carlo (CMC) to assess the microstructural properties of the wurtzite InGaN. The Gibbs free-energy was coarse grained as a function of parametric chemical potential (µ) and temperature (T) of the wurtzite $In_xGa_{1-x}N$, which was calculated by performing the Grand Canonical Monte Carlo (GCMC)[74] simulations using the Metropolis algorithm of CASM, and using a very large 16x16x16 supercell with 16,384 atoms for the bulk incoherent model and 24 x 24 x 7 with 16,128 atoms for the epitaxially coherent model. The simulation used the automatic convergence scheme where the formation energies ($E_f$) are converged within $10^{-3}$ eV/f.u. The hysteresis of the x vs. µ was removed using the thermodynamic integration, and phase boundaries were identified by analyzing the discontinuity in x vs $\mu$ curves as we did in our previous work[51].

The In compositional fluctuations and phase segregations at different T and x were investigated using the Canonical Monte Carlo performed on the 24x24x7 supercells with 16,128 atoms. Note, the shortest length of the supercell corresponds to the typical width (~ 3 nm) of the $In_xGa_{1-x}N$ QWs.

**Sample Fabrication:**

The LED devices were prepared based on the recipe by Zhang *et al.*[47]. The epitaxial fabrication was carried out using a MOCVD on an 8" Si (111) substrates in an AIXTRON CRIUS® close-coupled-showered (CCS) reactor. Prior to the growth native oxide on the Si substrates were removed by in-situ annealing. A buffer layer of step graded $Al_xGa_{1-x}N$ was used to account for the lattice mismatch between GaN and the Si substrate. Over, this AlGaN layer, Si-doped n-GaN region of 2.4 $\mu m$ is grown followed by the active LED region. The active region consists of 3 sets of QWs is discussed **in Figure 1**(a). Each set of layers consist of 3 InGaN QW regions and GaN barrier layers with thickness 2.5 nm and 10 nm, respectively. The In content in the QW regions were controlled by varying the deposition temperature. The low Indium content (18.4%), intermediate Indium content (24.2%) and high indium content (27%) QWs were grown at different temperatures 760 °C. 725 °C and 685 °C, respectively. Finally, a p-AlGaN electron blocking layer (EBL) is deposited to prevent the electron leakage to the Mg doped p-GaN region is deposited. TMAl, TMGa, TMIn, and $NH_3$ were used as the precursors for Al, Ga, In and N respectively. Furthermore, the growth process was carried out in $H_2$ ambient condition.

**Preparation of the TEM Samples:**

The electron transparent samples were prepared in the [11-20] zone axis by *in situ* lift-off method in FEI versa Focused Ion Beam/Scanning Electron Microscopy (FIB/SEM) dual beam machine[75]. Initially, a TEM lamella was prepared from InGaN LED bulk by creating two trenches on both side of the lamella followed by undercutting. The lamella was then transferred to a Cu grid using nanoprobe built in the FIB/SEM machine. Thus, the sample was thinned to ~100 nm using a 30 kV Ga⁺ beam and subsequent final cleaning step was carried out in 2 kV. Before the microscope analysis, the TEM lamella was cleaned with a plasma at ~300 V and for ~60 mins to remove any residual contamination. The final samples were ~ 40 nm thick as measured from the log-ratio EELS method. The thickness determined by this method is an upper bound to the actual sample thickness as the EELS log ratio method overestimates the sample thickness owing to the surface plasmon scattering[37,76].

**Aberration-Corrected EELS Acquisition and Processing:**

The EELS results were acquired using a JEOL JEM-ARM200F equipped with a cold field emission gun (CFEG) and a 5$^{th}$ order ASCOR aberration corrector operated at 80 kV. A convergence and collection semi-angle of ~31 mrad and ~77 mrad were used respectively for the acquisition of spectrums. A Gatan GIF quantum energy filter [77] is used to record the spectra using an energy dispersion of 0.4 eV/channel. The quantification of Ga was carried out using Ga *L*-edge (~1115 eV). The Ga signal was extracted using the elemental quantification toolkit of the GMS digital micrograph package [78,79].

The quantification of the compositional fluctuations was carried out using the Ga map using **Eq. 3**. The composition axis was determined using the pristine Ga region.

$$\frac{n_{Ga}^{QW}}{n_{Ga}^{pristine}} = \left[\frac{I_{Ga}^{QW}(\beta,\Delta)}{I_{Ga}^{pristine}(\beta,\Delta)}\right]\left[\frac{\sigma^{Ga}(\beta,\Delta)}{\sigma^{Ga}(\beta,\Delta)}\right] \qquad (3)$$

where $I$ represent the integrated spectrum, $β$ is the collection angle, $Δ$ is the integration width, and $σ$ the partial scattering cross-section. Since both regions of the Ga spectrum were used with same integration width, the scattering co-efficient of the cross- section at the numerator of Eq. 3 and the denominator cancel out. The pristine region was chosen at spacing of ~ 1 nm from the QW region to avoid the residual In gradient outside the QW during the growth of the InGaN LEDs.

The nominal In content is given by Eq. 4.

$$n_{In}^{QW} = 1 - \left[\frac{n_{Ga}^{QW}}{n_{Ga}^{pristine}}\right] \qquad (4)$$

The extraction of the QW region from the Ga map was carried out using an in-house developed MATLAB script.

**Estimating the Size of Compositional Fluctuations with the Laplacian of Gaussian:**

The sizes of the compositional fluctuations in the QW regions were determined using the autonomous scale-space method based on Laplacian of Gaussian (LoG)[44,80] implemented in the scikit-image[81] package. The smallest Gaussians used have a variance of ~2.3 Å —the typical length scale of In-N and Ga-N bonds, whereas the largest Gaussians have a variance of ~10.6 Å —the typical thickness of the QWs (~ 3 nm in diameter). A resolution of 9 Gaussians between 2.3 Å and 10.6 Å was used. A threshold of 0.10 times the average intensity value of the maps is given to eliminate the background noise, which can result in the erroneous determination of local maxima. The LoG kernel is parameterized with respect to the variance ($σ$) as given in **Eq. 4**.

$$\nabla^2 h(r) = -\left[\frac{r^2 - \sigma^2}{\sigma^4}\right] e^{-r^2/2\sigma} \tag{4}$$

where, *h* is a Gaussian with variance *σ* and *r* stands for real space coordinates of the image. The *σ* can be converted into the FWHM and Diameter of the compositional fluctuation using **Eq. 5** and **6**.

$$FWHM = 2\sqrt{2ln2}\,\sigma \tag{5}$$

$$Diameter = 2\sqrt{2}\,\sigma \tag{6}$$

The Laplacian of Gaussian can be modified to detect anisotropic clusters as shown is the S9 of the supplementary information. This has been implemented in the glog function in the scikit-image package which can be obtained from github. (https://github.com/caneparesearch/scikit-image/tree/glog).

**Scanning Transmission Electron Microscope-Cathodoluminescence Measurements:**

Cathodoluminescence (CL) imaging on the TEM foils were conducted on a polished sample using a 2010F field-emission TEM equipped with a Gatan MonoCL3 CL+ system operating at 80 kV at the temperature of liquid nitrogen. CL spectra were captured at 450 nm and 490 nm in the region of the QWs. The detailed setup is found in the study by Lim *et al*.[82].

**Incoherent Scattering analysis of the Monte Carlo Simulated Structures**

The Monte Carlo hexagonal structures were wrapped in the orthogonal cell with (1-100) and (11-20) bounding planes and rotated to the [11-20] zone axis using the atomsk software[83]. The final supercell compromised 16,128 atoms. The incoherent scattering approximation simulations were carried in the multi slice formalism with the isotropic frozen phonon approximation (10 phonon configurations) using the GPU based MULTEM simulation package[52,53]. The simulation parameters were fixed to the experimental parameters, using an accelerating voltage of 80 kV and a convergence angle of 31 mrad, which gives a depth of field of ~4.2 nm[84]. To approximate the incoherent scattering an ultra large detector at 100 mrad inner angle and 270 mrad outer angle were used. The real space sampling of 0.09 nm was used in the simulations to match the experimental resolution. To account for the source broadening a gaussian convolution with 0.2 nm was made in the final micrograph. At each temperature four Monte Carlo configurations were simulated each with a defocus value of -4.5 nm. Therefore, to obtain the proper statistics at each temperature and concentration a total of 15 Monte Carlo configurations have been analyzed.

**Acknowledgements**

This research was supported by the National Research Foundation (NRF) Singapore through Singapore MIT Alliance for Research and Technology (SMART)'s Low Energy Electronic Systems (LEES) IRG. P. C. acknowledges funding from the National Research Foundation under his NRF Fellowship NRFF12-2020-0012. Z. D. and P. C. are grateful to the Green Energy programme under the project code R284-000-185-731. The computational work was performed on resources of the National Supercomputing Centre, Singapore (https://www.nscc.sg) and NUS HPC (https://nusit.nus.edu.sg/hpc/). P. C. and T. P. M. acknowledge fruitful discussion with Prof. S. G. Gopalakrishnan at the Indian Institute of Science, Bangalore. T.P.M acknowledges insightful discussions with Dr. Changjian Li, National University of Singapore.

**Author Contribution**

TPM developed different aspects of the project under the supervision of SG, SJP and PC. TPM and ZD performed the computational modelling under the supervision of PC. TPM, ZD and PC performed the modelling data analysis. GJS and CJY prepared sample of electron microscopy studies. GJS and TPM obtained the EELS spectrum. SAG obtained the CL spectrums. LZ provided the InGaN QW sample. TPM developed the image processing algorithm. TPM conducted the EELS analysis and multislice simulation. TPM and PC put together the different aspects of the project and prepared the figures. TPM, SJP and PC drafted the first version of manuscript. All authors contributed to the scientific discussion and the final version of the manuscript.

# Unlocking the Origin of Compositional Fluctuations in InGaN Light Emitting Diodes —Supplementary Information—

Tara Prasad Mishra,[†,‡,∥] Govindo J. Syaranamual,[‡,∥] Zeyu Deng,[*,†] Jing Yang Chung,[†,‡] Li Zhang,[‡] Sarah A Goodman,[¶] Lewys Jones,[§] Michel Bosman,[†] Silvija Gradecak,[*,†,‡,¶] Stephen Pennycook,[*,†,‡] and Pieremanuele Canepa[*,†,‡]

†Department of Materials Science and Engineering, National University of Singapore, 9 Engineering Drive 1, 117575 Singapore , Singapore

‡Singapore-MIT Alliance for Research and Technology, 1 CREATE Way, #10-01 CREATE Tower, Singapore 138602, Singapore

¶Massachusetts Institute of Technology, Department of Materials Science and Engineering, Cambridge, Massachusetts, 02139, USA

§School of Physics, Trinity College Dublin, Dublin 2, D02 PN40, Ireland

∥Contributed equally to this work

E-mail: msedz@nus.edu.sg; gradecak@nus.edu.sg; steve.pennycook@nus.edu.sg; pcanepa@nus.edu.sg

# Contents

# S1 EELS Imaging Below Knock-on Threshold

Figure S1 validates that imaging under the ~80 kV accelerating voltage is non damaging the QWs. Even after a prolonged beam exposure (~10 min) the QWs are preserved without any damage to the structural integrity compared to fuzzy QWs, which are generally observed as a result of beam damage.[1] A general EELS measurement takes about ~1 min for the data acquisition, which suggests we had imaged the structure rather than beam artefacts.

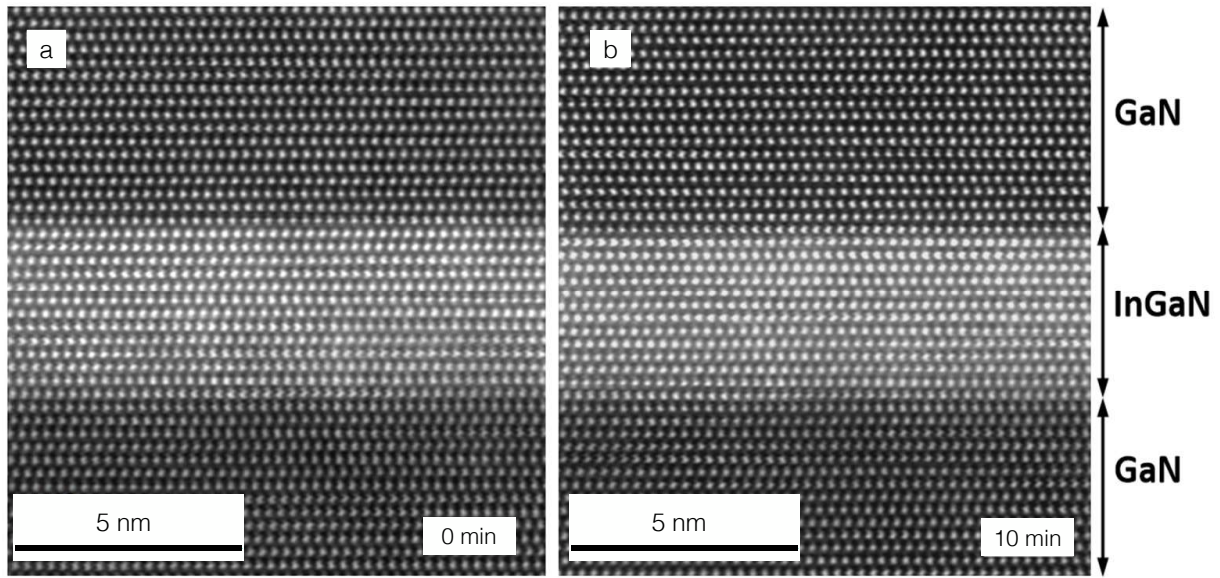

Figure S1: High angle annular dark field (HAADF) images taken (a) before and (b) after 10 min exposure to the electron beam at ~80 kV on the same LED device (see manuscript), but at a different spatial region from that used in the EELS imaging of the main manuscript.

# S2 Compositional Fluctuations Detected by EELS

The analyzed EELS compositional maps for the low In (∼ 18.4%) and high (∼24.2% In) are shown in columns A and B of Table S1, respectively.

Table S1: Heat maps of In composition in the QW regions of the InGaN LEDs as obtained from the EELS micrographs overlaid with the compositional fluctuations found using the Laplacian of Gaussian (LoG) algorithm for the blob detection. The blobs detected are marked as yellow circles. Each map is normalised with respect to the maximum and minimum In composition. The white bars set a spatial scale of 2 nm, while the vertical color bars indicate the In concentration.

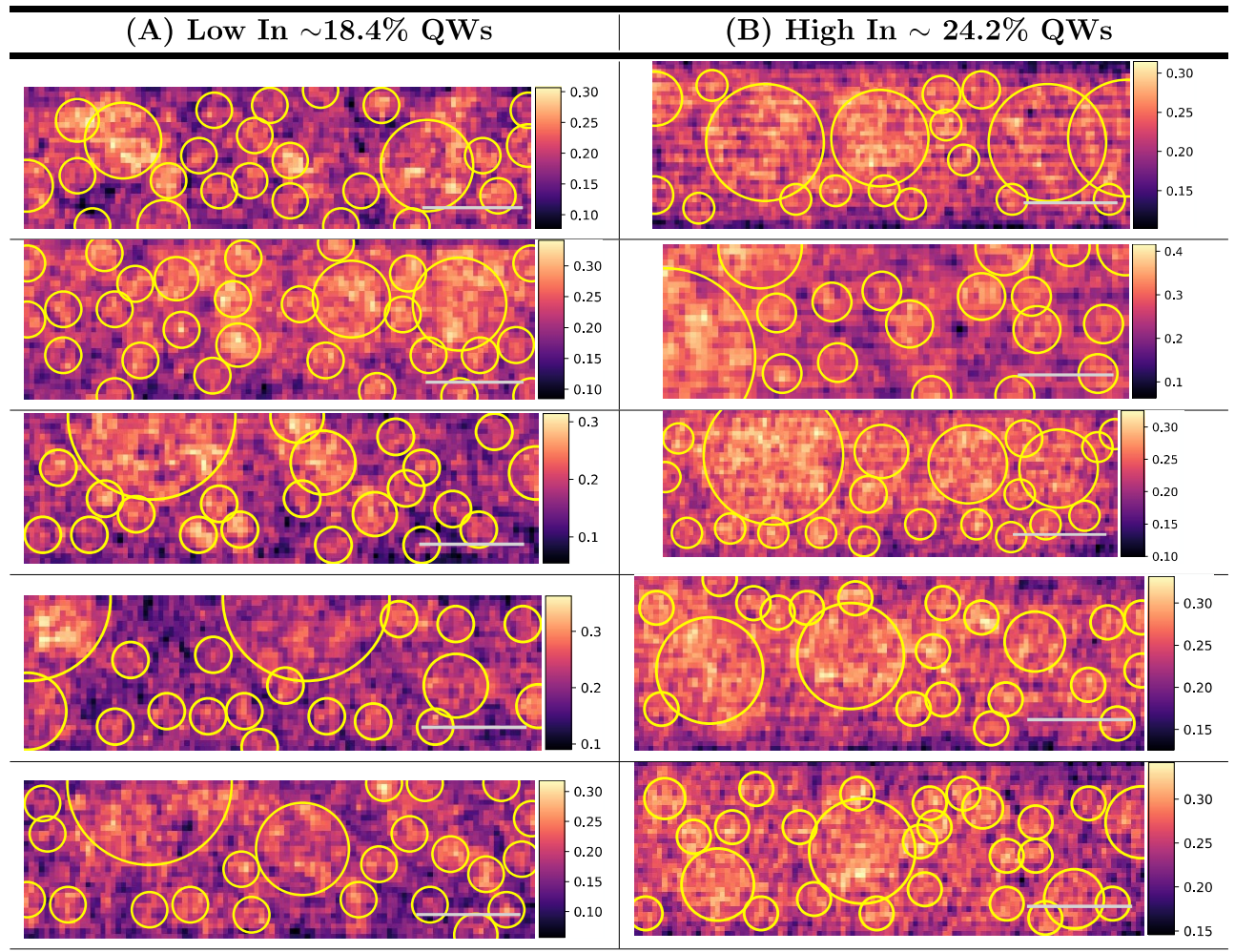

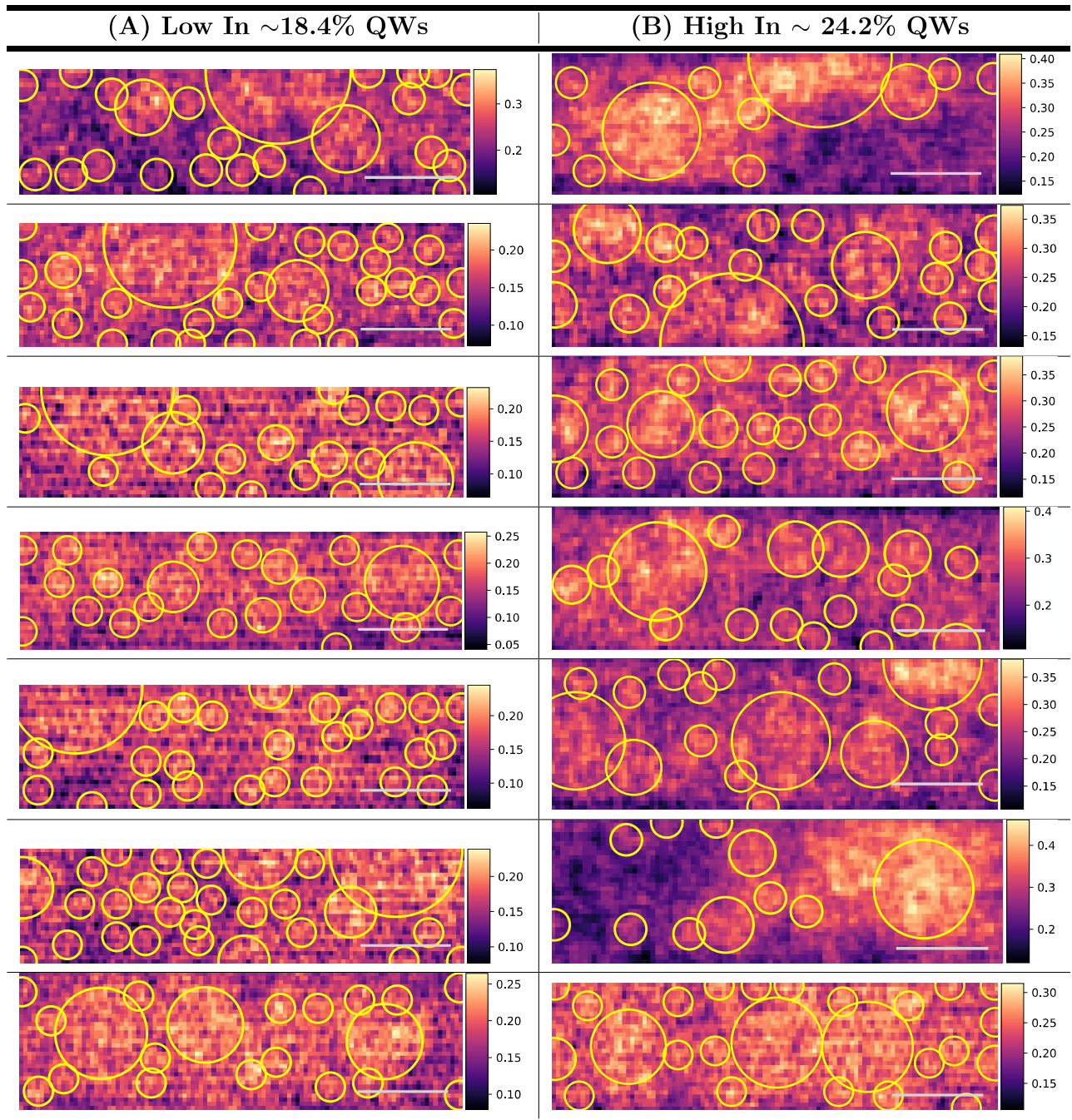

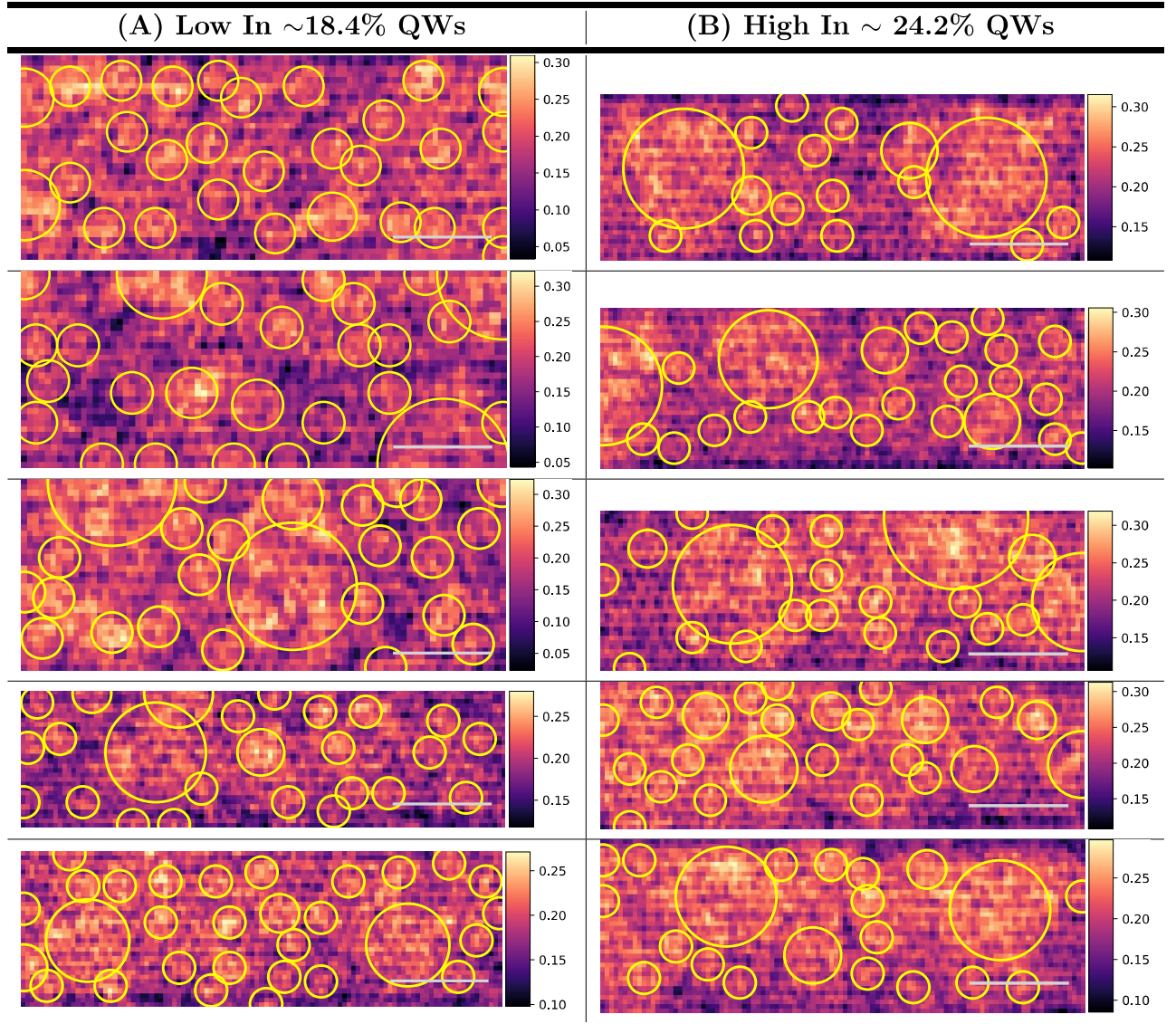

## S3  Multiscale Modeling of In$_x$Ga$_{1-x}$N System

A cluster expansion model as explained in the main text was used to efficiently coarse grain the Gibbs energy at variable chemical potential and temperature using the Grand Canonical Monte Carlo (GCMC) method for two different models of In$_x$Ga$_{1-x}$N QWs, namely bulk incoherent (fully relaxed in all direction) and epitaxially coherent (constrained to the GaN substrate and allowed to relax in the [0001] growth direction). Based on this approach we isolated 36 and 37 non-zero representative ECIs for the bulk incoherent and epitaxially coherent models, repsectively. The CEs including the ECIs are provided as JSON files which can be imported in the CASM software package.

Figure S2 plots the most significant ECIs (normalised by their multiplicity) as function of their index #.

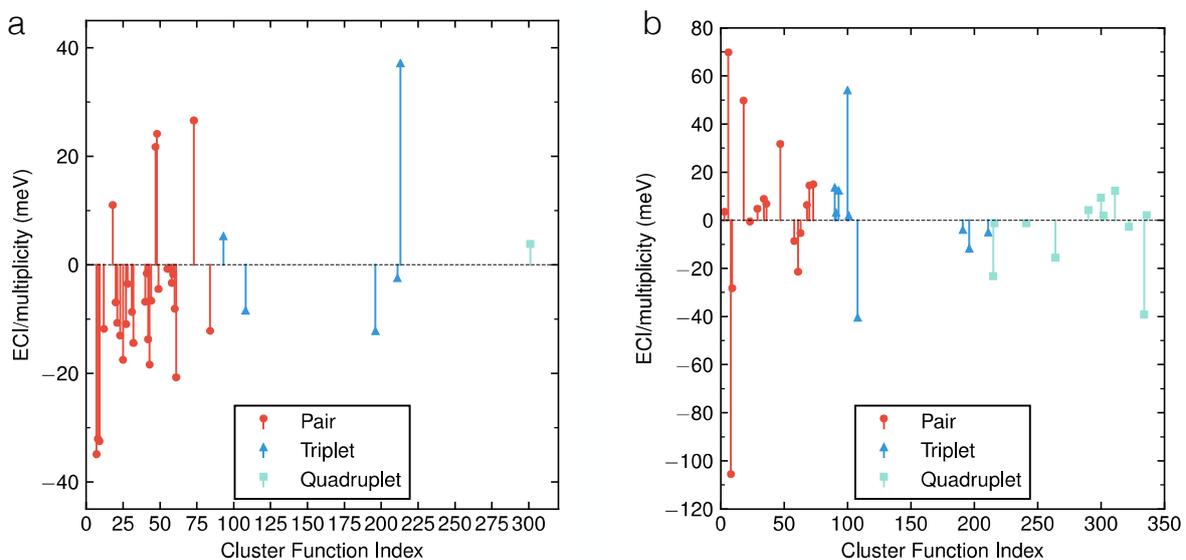

Figure S2: Normalized ECIs vs. their cluster index for (a) bulk incoherent and (b) epitaxially coherent models. The point terms are not shown.

In the case of the bulk incoherent model, out of the 36 ECIs, 29 ECIs are pair interactions, 5 are triplets and 1 is a quadruplet. Figure S2(a) shows that the most dominant and stabilising ECIs are represented by pair interactions, with cluster # 7 being the most stable normalized ECI (−34.902 meV) and expanding a short length ($\sim$ 4.506 Å). Other pair interactions identified by clusters # 8 (−32.084 meV) and # 9 (−32.498 meV) contribute significantly in lowering the formation energy. Expectedly, triplet and quadruplet ECIs contribute to the CE but extend to smaller radii compared to pair ECIs. The triplet # 213 (37.129 meV) has a destabilizing contribution to the CE.

For the epitaxially coherent model, out of the 37 ECIs, 16 ECIs are pair interactions, 9 are triplets and 11 are quadruplet terms as shown in Figure S2(b). The addition of strain, constraining the lattice to the GaN substrate, results in an increase of the number of stabilizing ECIs, i.e. ECIs with negative values. Similar to the bulk incoherent system the pair interaction # 8 still form the dominant stabilizing ECI (−105.515 meV) and expanding at short length (∼5.185 Å). Other pair interactions identified by clusters # 9 (−28.249 meV) and # 61 (−21.404 meV) contribute significantly in lowering the formation energy. The pair interactions # 6 (69.819 meV) and # 18 (49.757 meV) together with the # 100 (53.967 meV) have a major destabilizing effect. When turning the attention to triplet interactions, ECI # 108 (−40.514 meV) has a stabilizing effect and extends to a smaller radii of ∼ 3.181 Å. Interestingly, for the epitaxially coherent case a large number of short range stabilizing quadruplets are observed which are absent in the bulk incoherent case. These stabilizing quadruplet clusters are # 215 (−23.269 meV), # 264 (−15.529 meV) and # 334 (−39.135 meV), which all extend to small lengths of ∼3.19 Å. The increase in large number of stabilizing ECIs due to addition of strain results in the creation of monophasic regions, as depicted in the epitaxially coherent phase diagram shown in Figure 4(b) of the main text.

# S4 Accuracy of the Cluster Expansion Models

Figure S3 shows the error obtained from the cluster expansion (CE) models benchmarked on the DFT data for the bulk incoherent and epitaxially coherent conditions.

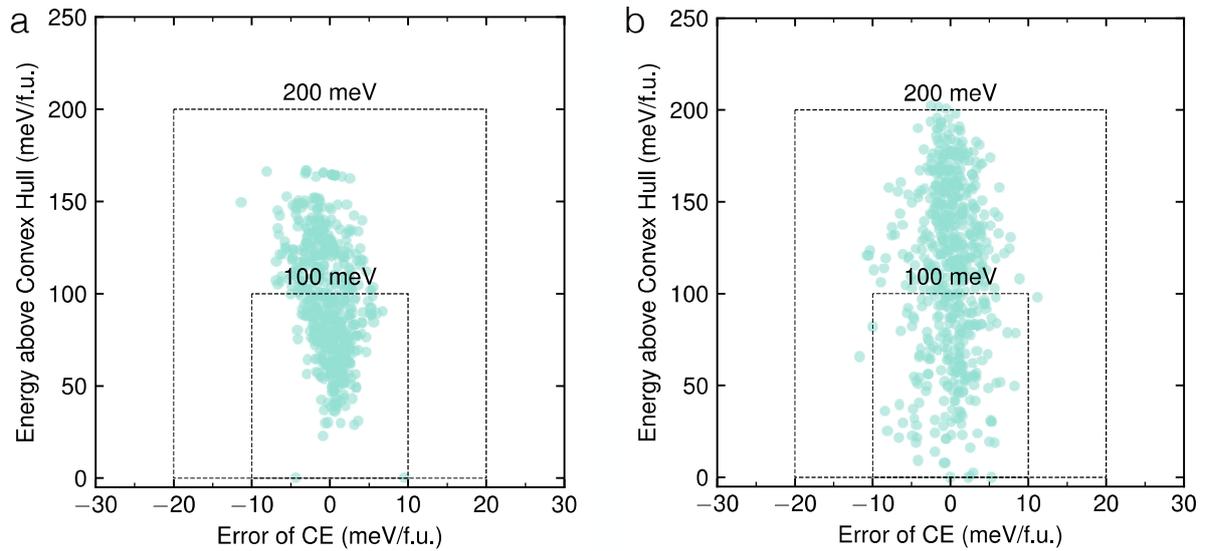

Figure S3: Error obtained from CE benchmarked on the DFT data for InGaN QWs in (a) bulk incoherent model and panel (b) for the epitaxially coherent model.

The CEs model developed are able to reproduce all the DFT calculated formation energy with high accuracy. The maximum error for structures in the convex hull is consistently below 10 meV/f.u.

## S5 Stable Orderings in Epitaxially Coherent InGaN QWs

Addition of strain stabilizes intermediate monophasic superlattice structures at In concentration of $x \sim 0.33$ and $x \sim 0.50$ and as shown in Figure S4. The structure at $x \sim 0.33$ is a metastable structure with a small energy above the convex hull ($\sim 1.43$ meV/f.u.). This structure is formed beyond the temperature $\sim 100$ K as seen in the phase diagram in figure 4 of the main text. The structure at $x \sim 0.50$ is a configuration which lies in the convex hull.

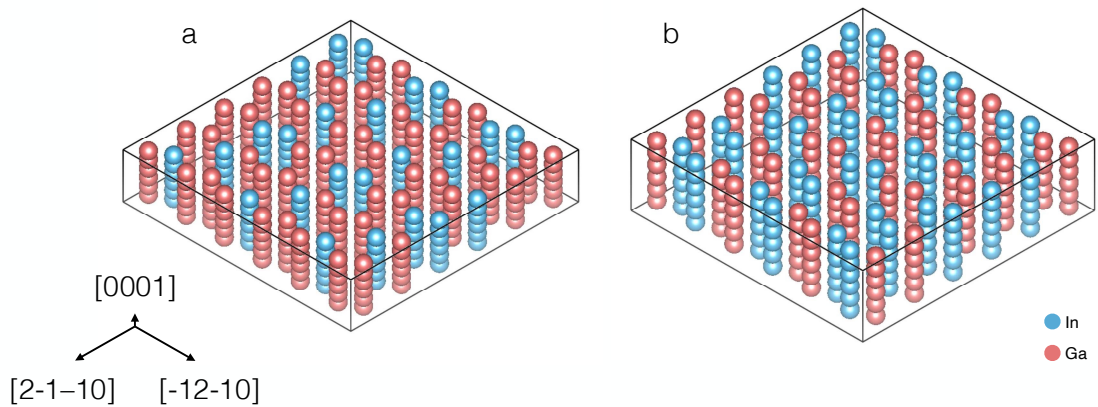

Figure S4: (a) Metastable structure ($\sim 1.43$ meV f.u.$^{-1}$ above the convex hull) $(InN)_1/(GaN)_3$ at $x = 0.33$. (b) Ground state structure, $(InN)_2/(GaN)_2$, of a $In_xGa_{1-x}N$ wurtzite alloy at $x = 0.50$. These structures are stabilized due to the epitaxial constrain to the GaN substrate.

# S6    Thermodynamic Integration

The grand-canonical Monte Carlo (GCMC) calculations were performed defining the grand-canonical potential $\Phi$ as in Eq. 1.

$$\Phi = [E - TS] - \mu x; \tag{1}$$

where $E$ is the energy calculated from cluster expansion, $S$ is the configurational entropy, $T$ is the temperature, and $\mu$ is the chemical potential of In and $x$ is the composition in $In_xGa_{1-x}N$.

$\Phi$ is evaluated through a thermodynamic integration:

$$\Phi(\beta, \mu) = \frac{\beta_0}{\beta}\Phi_0(\beta_0, \mu) + \frac{1}{\beta}\int_{\beta_0}^{\beta}[E - \mu x]\,d\beta; \tag{2}$$

$$\text{with } \Phi_0(\beta_0, \mu) = E - \mu_{In}x. \tag{3}$$

where $\beta = \frac{1}{k_BT}$ and $k_B$ is the Boltzmann constant.

The integration starts from the $T = 10$ K and up to 2500 K with a step of $\Delta T = 10$ K in a $\mu_{In}$ range of $\pm$ 0.325 eV/f.u. Then at each $T$, $\mu_{In}$ was scanned in both forward (from $\mu = -1.5$ to 1.5 eV) and backward (from $\mu = 1.5$ to $-1.5$ eV) directions with a step of $\Delta\mu = 0.05$ meV, and $\Phi$ was integrated using:

$$\Phi(\beta, \mu_{In}) = \Phi_0(\beta, \mu_{0,In}) - \frac{1}{\beta}\int_{\mu_0}^{\mu} x\,d\mu_{In}; \tag{4}$$

$$\text{with } \Phi_0(\beta, \mu_{0,In}) = \Phi_{\text{heating}}(\beta, \mu_{0,In}.) \tag{5}$$

Due to the negligible effects of entropy at low temperatures (near 0 K), the starting values of $\Phi_0$ can be approximated as $\Phi_0 = E - \mu_{In}x$. The integration at each $\mu_{In}$ starts from the $\Phi_{heating}(T, \mu_0)$ where $\mu_{0,In} = -0.01$ eV/f.u.

# S7 Representative Snapshots and EELS Maps from CMC

The canonical Monte Carlo (CMC) snapshots at each temperature are presented in Table S2 and S3. To highlight the variation in the distribution of In atoms in the model supercells, only In atoms (in red) are shown in the left column of Table S2 and S3.

The hexagonal structures generated by CMC were wrapped in an orthogonal cell with $(1-100)$ and $(11-20)$ bounding planes and rotated to the $[11-20]$ zone axis using the atomsk software.[2] The incoherent scattering simulations were carried out in the multislice formalism using the GPU based MULTEM simulation package.[3,4]

In order to perform a statistical analysis of the compositional fluctuation at each condition of temperature and In concentration 15 different CMC simulations were included. To sample through the CMC model supercell of each configuration, multislice simulations at defocus value of –40.0 Å, —probe position along the beam direction— were performed. The micrograph generated were then used to quantitatively extract the compositional fluctuations using the blob detection using the Laplacian of Gaussian (LoG) algorithm described in the main text.

Table S2: Column (a) CMC hexagonal structures generated at given temperatures and In concentrations for bulk incoherent conditions. Column (b) Multislice simulations of CMC structures along the $[11-20]$ zone axis after transforming the hexagonal cell to orthogonal supercells bounded by the $(11-20)$ and $(1-100)$ family of planes, respectively. The white bar sets a 2 nm scale, while the color bar indicates In-rich and In-poor areas, respectively.

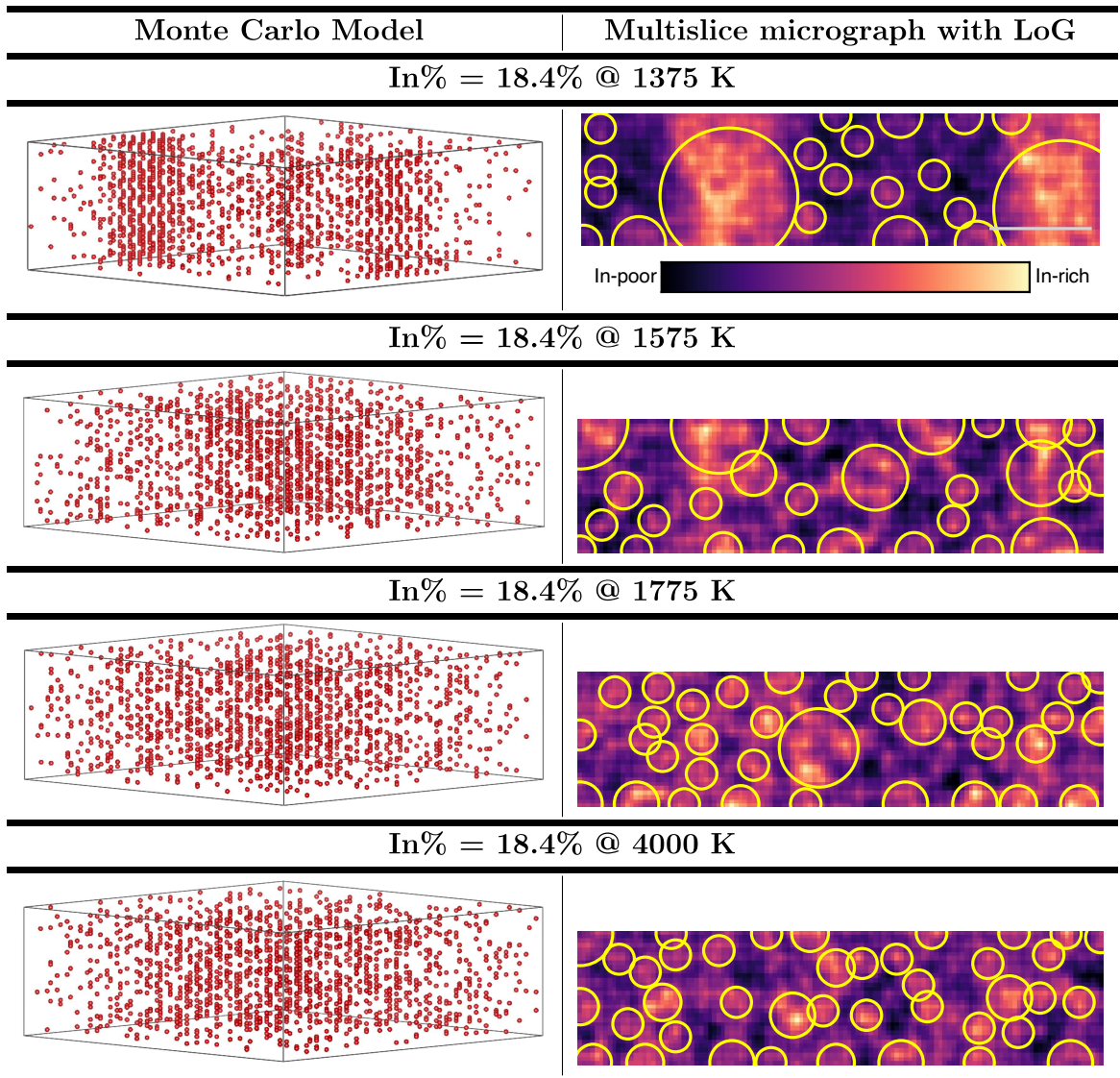

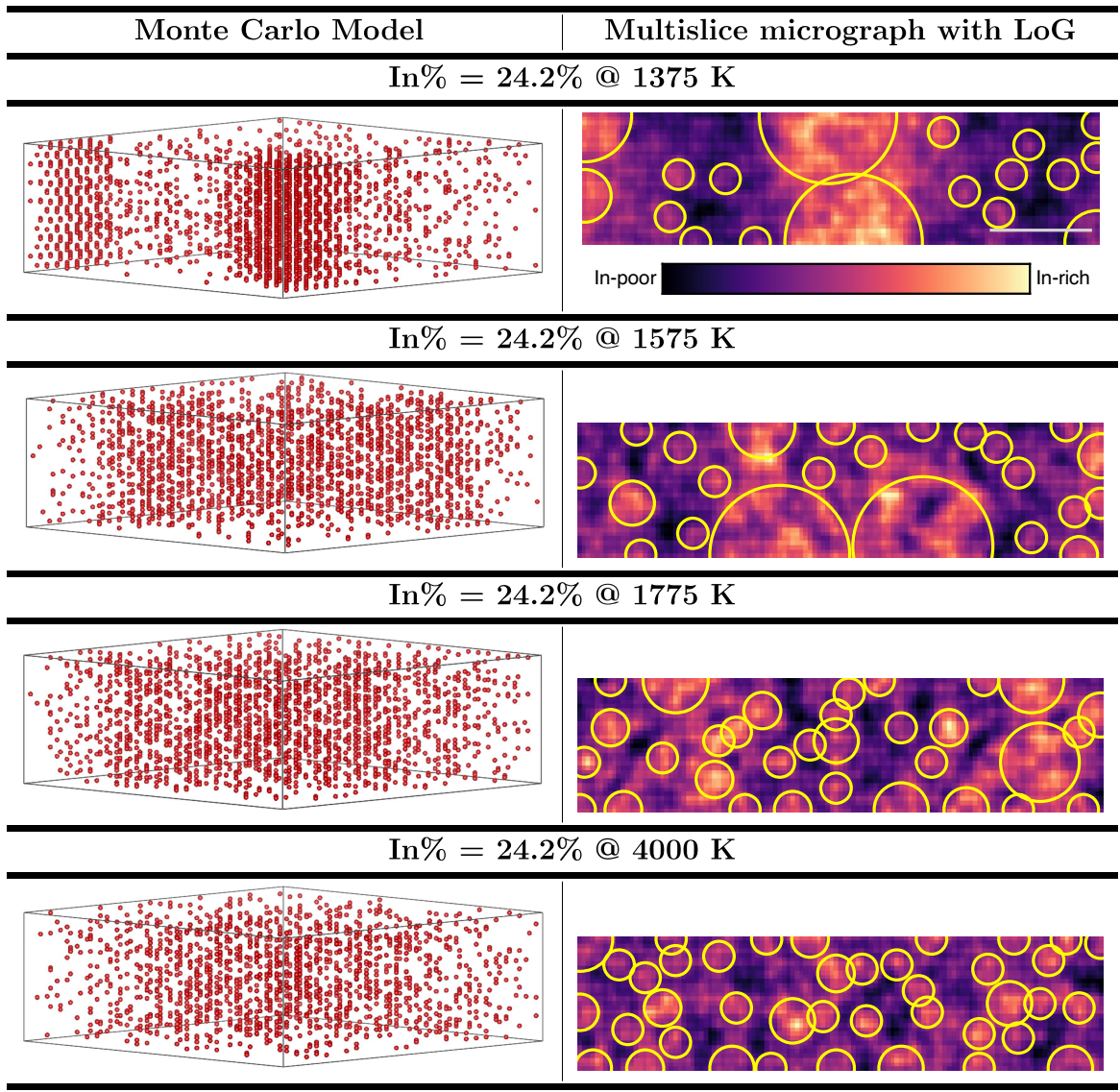

Table S3: Column (a) CMC hexagonal structures generated at given temperatures and In concentrations for epitaxially coherent conditions. Column (b) Multislice simulations of CMC structures along the $[11-20]$ zone axis after transforming the hexagonal cell to orthogonal supercells bounded by the $(11-20)$ and $(1-100)$ family of planes, respectively. The white bar sets a 2 nm scale, while the color bar indicates In-rich and In-poor areas, respectively.

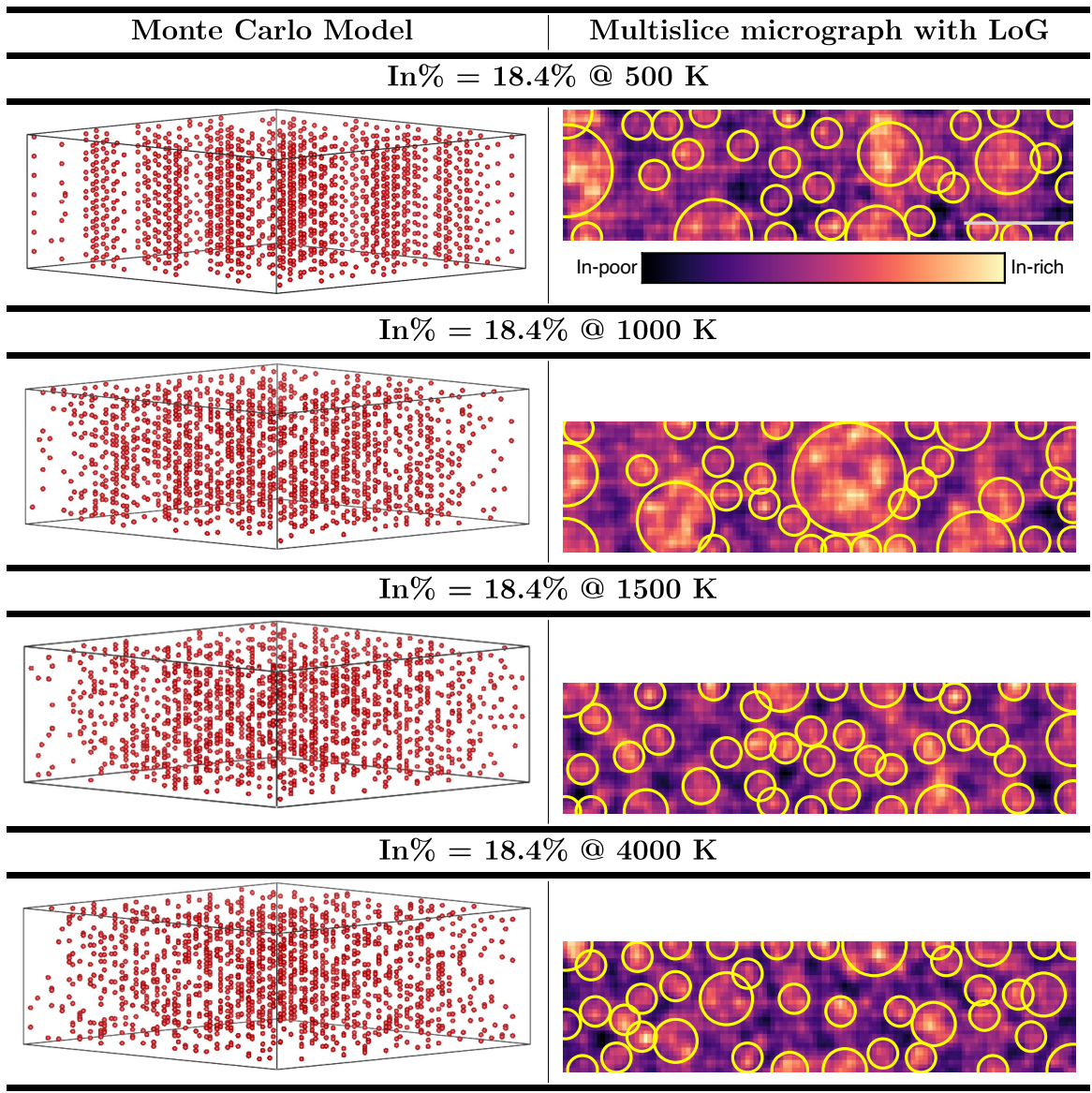

| Monte Carlo Model | Multislice micrograph with LoG |
|---|---|
| **In% = 24.2% @ 500 K** ||
| 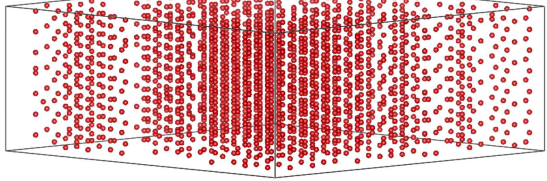 | 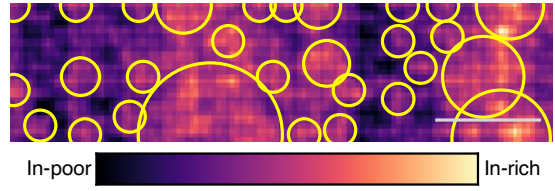 |
| **In% = 24.2% @ 1000 K** ||
| 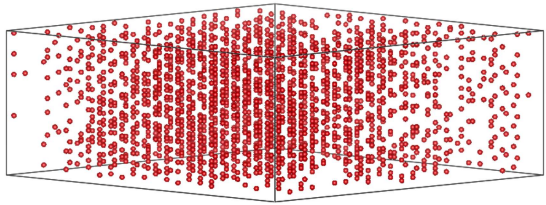 | 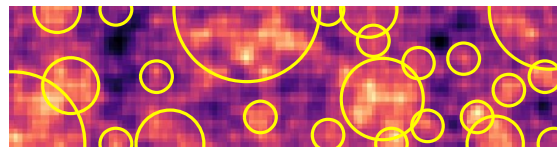 |
| **In% = 24.2% @ 1500 K** ||
| 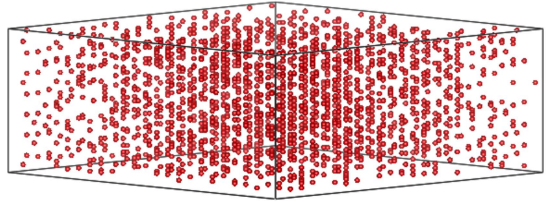 | 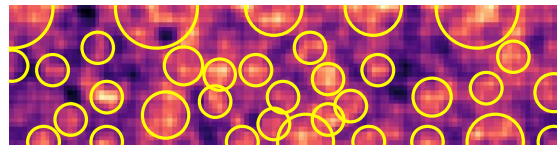 |
| **In% = 24.2% @ 4000 K** ||
| 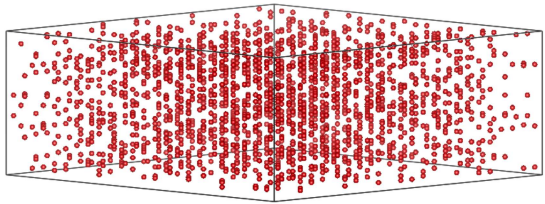 | 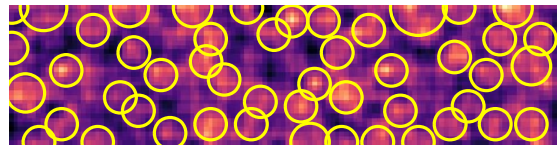 |

# S8  Compositional Fluctuations vs. Temperature

Figure S5 shows the variation of distribution of compositional fluctuations by changing the temperature of the bulk incoherent CMC simulations using a violin-type reppresentation.

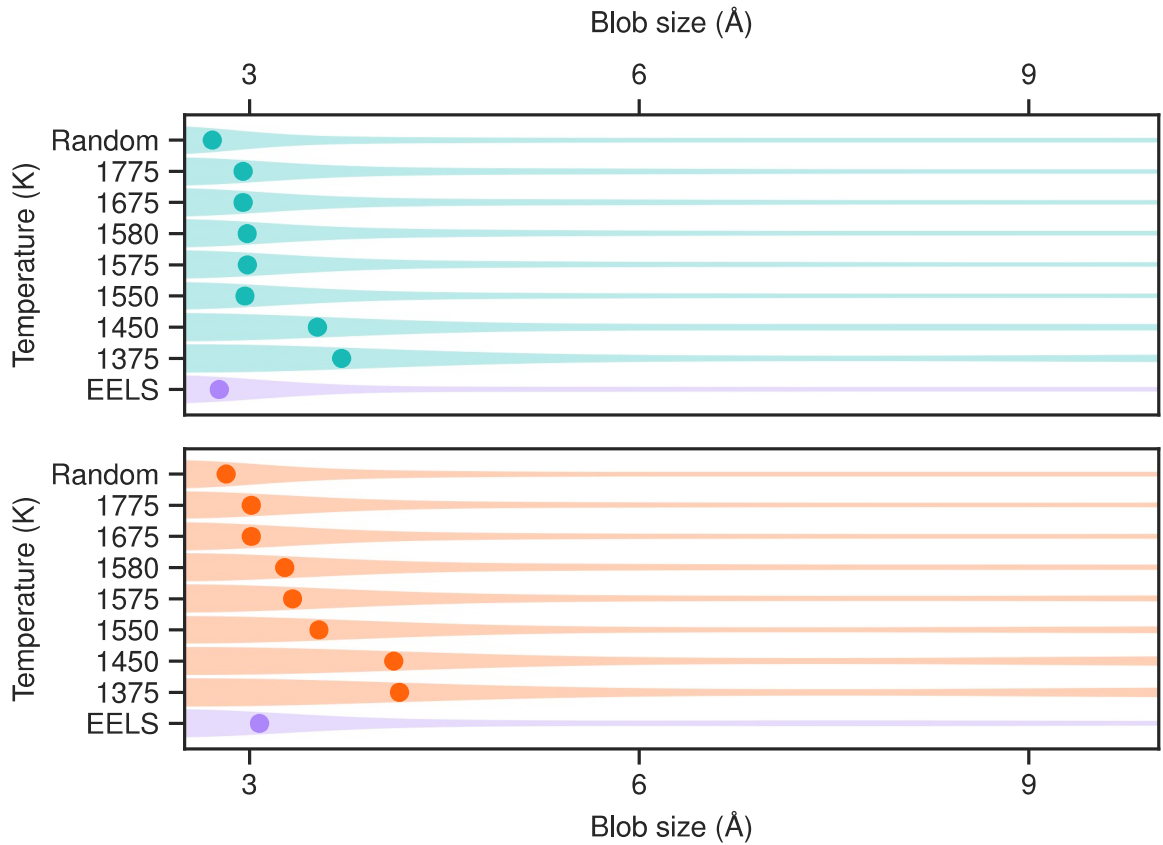

Figure S5: The top panel shows the distribution of the compositional fluctuations size for samples with In concentration of ∼18.2% and ∼24.2% in bottom panel. Where indicated the micrographs of the QWs are obtained from EELS experiments or alternatively from the CMC simulations at specific temperatures fir bulk incoherent model. The random alloy distribution is obtained by running the CMC simulations at a temperature (∼4,000 K) significantly higher than the miscibility temperature.

The x-axis depicts the size of potential In compositional fluctuations revealed experimentally (purple violin shapes) or theoretically predicted at different temperatures 1375, 1450, 1550, 1575, 1580, 1675, 1775 and 4000 K (green and orange violins), respectively. The vertical width of each violin plot is the probability density of an In compositional fluctuation to exist with a specific cluster size. The average sizes of the compositional fluctuations obtained from the EELS and CMC calculations are marked by the circles in Figure S5.

Figure S5 shows the variation of distribution of compositional fluctuations by changing the temperature of the epitaxially coherent CMC simulations using a violin-type reppresentation. Figure S6 shows the variation of distribution of compositional fluctuations by changing the temperature of the epitaxially coherent CMC simualtions using a violin-type reppresentation.

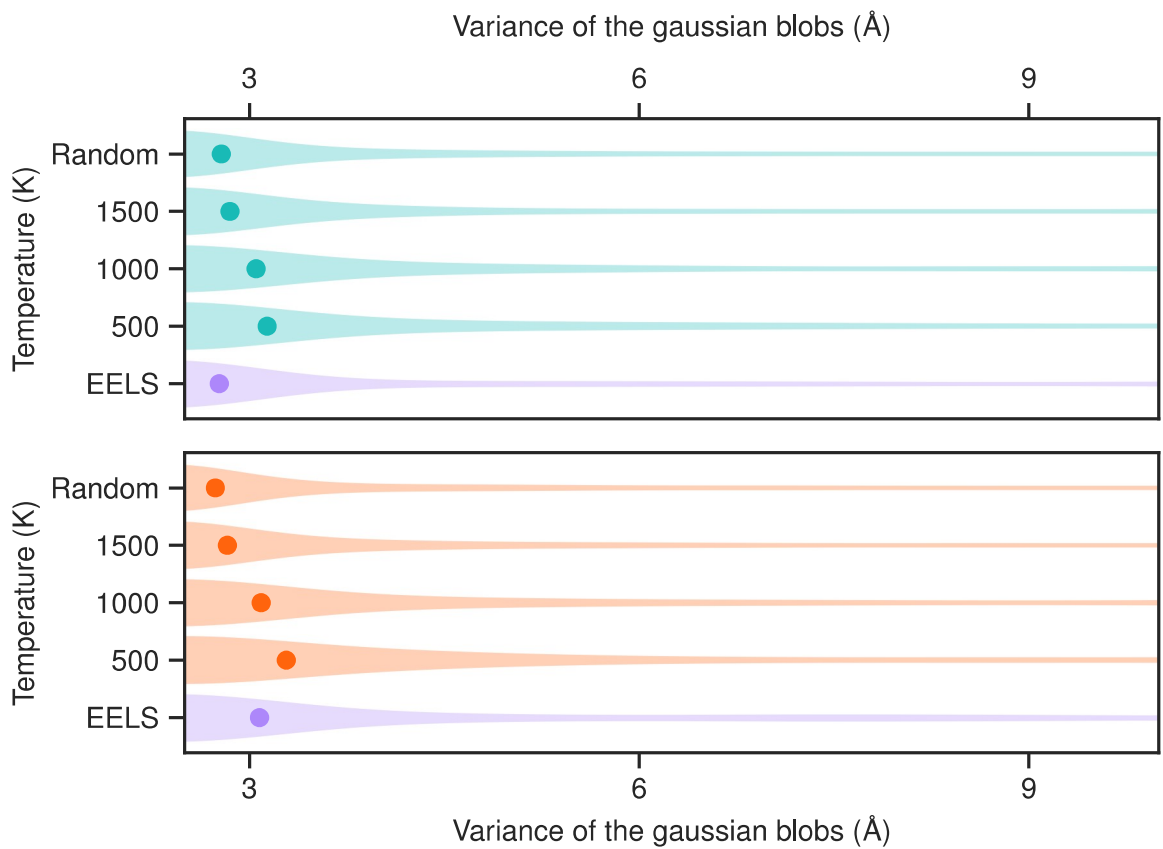

Figure S6: The top panel shows the distribution of the compositional fluctuations size for samples with In concentration of ∼18.2% and ∼24.2% in bottom panel. Where indicated the micrographs of the QWs are obtained from EELS experiments or alternatively from the CMC simulations at specific temperatures. The random alloy distribution is obtained by running the CMC simulations at a temperature (∼4,000 K) significantly higher than the miscibility temperature.

The x-axis depicts the size of potential In compositional fluctuations revealed experimentally

(purple violin shapes) or theoretically predicted at different temperatures 500, 1000, 1500 and 4000 K (green and orange violins), respectively. The vertical width of each violin plot is the probability density of an In compositional fluctuation to exist with a specific cluster size. The average sizes of the compositional fluctuations obtained from the EELS and CMC calculations are marked by the circles in Figure S6.

# S9 Anisotropic Compositional Fluctuations Detected using Laplacian of Gaussian

The Laplacian of Gaussian algorithm can be extended to include anisotropic compositional fluctuations in the InGaN QWs. The analyzed EELS compositional maps for the low In ($\sim 18.4\%$) and high ($\sim 24.2\%$ In) are shown in columns A and B of Table S4, respectively. The compositional fluctuations are detected using a different value of $\sigma$ in the X and Y direction. For the sake of simplicity we have approximated the compositional fluctuations as circular Gaussians in our current work. The anisotropic Laplacian of Gaussian has been implemented using the developed function glog in the scikit-image package. The package can be accessed from github(https://github.com/caneparesearch/scikit-image/tree/glog).

Table S4: Heat maps of In composition in the QW regions of the InGaN LEDs as obtained from the EELS micrographs overlaid with the anistropic compositional fluctuations found using the Laplacian of Gaussian (LoG) algorithm for the blob detection. The blobs detected are marked as yellow circles. Each map is normalised with respect to the maximum and minimum In composition. The white bars set a spatial scale of 2 nm, while the vertical color bars indicate the In concentration.

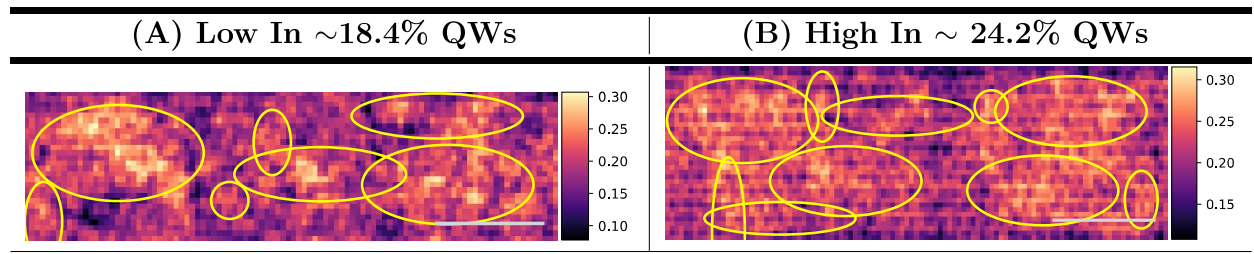

| (A) Low In ~18.4% QWs | (B) High In ~ 24.2% QWs |
| --- | --- |